\documentclass[twocolumn]{emulateapj}

\newcommand{\msec}{\,{\rm m}\,{\rm s}^{-1}}	
\newcommand{\kms}{\,{\rm km}\,{\rm s}^{-1}}	
\newcommand{\vunits}{\,h^3\,{\rm Mpc}^{-3}}
\newcommand{\mangstrom}{\,{\rm \AA}}
\newcommand{\cm}{\,{\rm cm}}
\newcommand{\teff}{T_{\rm eff}}
\newcommand{\K}{\,{\rm K}}
\newcommand{\sn}{{\rm S/N}}
\newcommand{\hmpc}{\,h^{-1}{\rm Mpc}}

\newcommand{\lya}{Ly$\alpha$}		
\newcommand{\feh}{[Fe/H]}			

\newcommand{\snr}{{\rm S/N}}

\slugcomment{Preprint Version, 1/10/11}

\shorttitle{SDSS-III}
\shortauthors{Eisenstein et al.}

\begin{document}


\title{SDSS-III: Massive Spectroscopic Surveys of the 
  Distant Universe, the Milky Way Galaxy, and Extra-Solar
  Planetary Systems}

\author{
Daniel J. Eisenstein\altaffilmark{1,2},
David H. Weinberg\altaffilmark{3,4},
Eric Agol\altaffilmark{5},
Hiroaki Aihara\altaffilmark{6},
Carlos Allende Prieto\altaffilmark{7,8},
Scott F. Anderson\altaffilmark{5},
James A. Arns\altaffilmark{9},
\'Eric Aubourg\altaffilmark{10,11},
Stephen Bailey\altaffilmark{12},
Eduardo Balbinot\altaffilmark{13,14}
% ...
}

\altaffiltext{1}{
Steward Observatory, 933 North Cherry Avenue, Tucson, AZ 85721, USA
}

\altaffiltext{2}{
Harvard College Observatory, 60 Garden St.,
Cambridge, MA 02138, USA
}

\altaffiltext{3}{
Department of Astronomy, 
Ohio State University, Columbus, OH 43210, USA
}

\altaffiltext{4}{
Center for Cosmology and Astro-Particle Physics,
Ohio State University, Columbus, OH 43210, USA
}

\altaffiltext{5}{
Department of Astronomy, University of Washington, Box 351580, Seattle, WA
98195, USA
}

\altaffiltext{6}{
Institute for the Physics and Mathematics of the Universe,
The University of Tokyo,
5-1-5 Kashiwanoha, Kashiwa, 277-8583, Japan
}

\altaffiltext{7}{
Instituto de Astrof\'\i{}sica de Canarias, E38205 La Laguna, Tenerife, Spain
}

\altaffiltext{8}{
Departamento de Astrof\'{\i}sica, Universidad de La Laguna, 38206, La 
Laguna, Tenerife, Spain
}

\altaffiltext{9}{
Kaiser Optical Systems, 
371 Parkland Plaza
Ann Arbor, MI 48103, USA
}

\altaffiltext{10}{
Astroparticule et Cosmologie (APC),
Universit\'e Paris-Diderot,
10 rue Alice Domon et L\'eonie Duquet,
75205 Paris Cedex 13, France
}

\altaffiltext{11}{
CEA, Centre de Saclay, Irfu/SPP,  F-91191 Gif-sur-Yvette, France
}

\altaffiltext{12}{
Lawrence Berkeley National Laboratory, One Cyclotron Road,
Berkeley, CA 94720, USA
}

\altaffiltext{13}{
Instituto de F\'\i sica, UFRGS, Caixa Postal 15051, Porto Alegre, RS -  
91501-970, Brazil
}

\altaffiltext{14}{
Laborat\'orio Interinstitucional de e-Astronomia, - LIneA, Rua Gal.  
Jos\'e Cristino 77, Rio de Janeiro, RJ - 20921-400, Brazil
}

% ...


\begin{abstract}
Building on the legacy of the Sloan Digital Sky Survey (SDSS-I and II),
SDSS-III is a program of four spectroscopic surveys on three scientific themes:
dark energy and cosmological parameters, the history and structure of the
Milky Way, and the population of giant planets around other stars. 
In keeping 
with SDSS tradition, SDSS-III will provide regular public releases of all its 
data, beginning with SDSS Data Release 8 (DR8), which was made public
in January 2011 and includes SDSS-I and SDSS-II images and spectra
reprocessed with the latest pipelines and calibrations produced for
the SDSS-III investigations.
This paper presents an overview of the four surveys that comprise SDSS-III.
The Baryon Oscillation 
Spectroscopic Survey (BOSS) will measure redshifts of 1.5 million massive
galaxies and \lya\ forest spectra of 150,000 quasars, using the baryon 
acoustic oscillation feature of large scale structure to obtain
percent-level determinations of the distance scale and Hubble expansion rate
at $z<0.7$ and at $z\approx 2.5$. 
SEGUE-2, an already completed SDSS-III survey that is the 
continuation of the SDSS-II Sloan 
Extension for Galactic Understanding and Exploration (SEGUE),
measured medium-resolution ($R=
\lambda/\Delta\lambda\approx 1800$) optical spectra of 118,000 stars in a 
variety of target categories, probing chemical evolution, stellar kinematics 
and substructure, and the mass profile of the dark matter halo from the solar 
neighborhood to distances of 100 kpc.  APOGEE, the Apache Point Observatory 
Galactic Evolution Experiment, will obtain high-resolution ($R\approx 30,000$), 
high signal-to-noise ratio ($\sn\geq 100$ per resolution element), $H$-band 
($1.51\micron<\lambda<1.70\micron$) spectra of $10^5$ evolved, late-type
stars, measuring separate abundances for $\sim 15$ elements per star
and creating the first high-precision spectroscopic survey of {\it all} 
Galactic stellar populations (bulge, bar, disks, halo) with a uniform set of
stellar tracers and spectral diagnostics. The Multi-object APO Radial Velocity
Exoplanet
Large-area Survey (MARVELS) will monitor radial velocities of more than 8000
FGK stars
with the sensitivity and cadence ($10-40\msec$, $\sim 24$ visits per star)
needed to detect giant planets with periods up to two years, providing an 
unprecedented data set for understanding the formation and dynamical evolution 
of giant planet systems. As of January 2011, SDSS-III has obtained spectra of 
more than 240,000 galaxies, 29,000 $z\geq 2.2$ quasars, and 140,000 stars, 
including 74,000 velocity measurements of 2580 stars for MARVELS. 
\end{abstract}






\section{Introduction}
\label{sec:intro}

The Sloan Digital Sky Survey (SDSS; \citealt{york00}) and 
the Legacy Survey of SDSS-II performed deep imaging of
8400 deg$^2$ of high Galactic latitude sky 
in five optical bands, repeat imaging 
of an equatorial stripe in the southern Galactic cap
(roughly 25 epochs on 300 deg$^2$),
and spectroscopy of more than 900,000 galaxies,
100,000 quasars, and 200,000 stars \citep{abazajian09}.
In addition to completing the original SDSS goals, SDSS-II
(which operated from 2005-2008) executed a supernova survey
in the southern equatorial stripe \citep{frieman08b},
discovering more than 500 spectroscopically confirmed Type Ia
supernovae in the redshift range $0.1 < z < 0.4$, and 
it also performed an imaging
and spectroscopic survey of the Galaxy, known as SEGUE
(the Sloan Extension for Galactic Understanding and Exploration;
\citealt{yanny09}), with 3200 deg$^2$ of additional imaging and
spectra of 240,000 stars selected in a variety of target categories.
These surveys were accomplished using a dedicated 2.5-m
telescope\footnote{The Sloan Foundation 2.5-m Telescope
at Apache Point Observatory (APO), in Sunspot, NM.} 
with a wide field of view ($7\,$deg$^2$, $3^\circ$-diameter;
\citealt{gunn06}),
a large mosaic CCD camera \citep{gunn98}, a pair of
double spectrographs, each fed by 320 optical fibers plugged into
custom-drilled aluminum plates,
and an extensive network of data reduction and calibration
pipelines and data archiving systems.
The resulting data sets have supported an enormous range of 
investigations, making the SDSS one of the most influential
astronomical projects of recent decades
\citep{madrid06,madrid09}.

The achievements of SDSS-I and II and the exceptional power
of the SDSS facilities for wide-field spectroscopy
together inspired SDSS-III, a six-year program begun in July 2008
and consisting of four large spectroscopic
surveys on three scientific themes: dark energy and cosmological
parameters, the history and structure of the Milky Way,
and the population of giant planets around other stars.
This paper provides an overview of the four
SDSS-III surveys, each of which will be described in greater
depth by one or more future publications covering survey 
strategy, instrumentation, and data reduction software.

The Baryon Oscillation Spectroscopic Survey (BOSS) is the
primary dark-time survey of SDSS-III.
It aims to determine the expansion history of the universe
with high precision by using the baryon acoustic oscillation
(BAO) feature in large-scale structure 
as a standard ruler for
measuring cosmological distances
\citep{eisenstein98,blake03,seo03}.
More specifically, the BOSS redshift survey of 1.5 million massive
galaxies aims to measure the distance-redshift
relation $d_A(z)$ and the Hubble parameter
$H(z)$ with percent-level precision out to $z=0.7$, using
the well established techniques that led to the first detections
of the BAO feature
\citep{cole05,eisenstein05}.
Pioneering a new method of BAO measurement,
BOSS will devote 20\% of its fibers to obtaining \lya\ forest
absorption spectra of 150,000 distant quasars,
achieving the first precision measurements of cosmic expansion
at high redshift ($z\approx 2.5$) and serving as a pathfinder
for future surveys employing this technique.
BOSS is also performing spectroscopic
surveys of approximately 75,000 ancillary science targets in
a variety of categories.  To enable BOSS to cover 10,000 deg$^2$,
the SDSS imaging camera was used at the start of SDSS-III to survey an
additional 2500 deg$^2$ of high-latitude sky in the 
southern Galactic cap; this imaging was completed in January 2010.
Because BOSS was designed to observe targets $1-2$ magnitudes fainter than
the original SDSS spectroscopic targets, 
substantial upgrades to the SDSS spectrographs were required.  
The upgraded spectrographs were commissioned 
in Fall 2009.  As of early January 2011, BOSS had obtained 
240,000 galaxy spectra and 29,000 high-redshift ($z\geq 2.2$) quasar spectra.

From July 2008 to July 2009, SDSS-III undertook a spectroscopic
survey of 118,000 stars in a variety of target categories,
using the original SDSS spectrographs.  This survey, called
SEGUE-2, is similar in design to the SEGUE-1 spectroscopic survey
of SDSS-II, but it used the results of SEGUE-1 to refine its
target selection algorithms.\footnote{We will henceforth use
the retrospective term ``SEGUE-1'' to refer to the SEGUE survey 
conducted in SDSS-II, and we will use ``SEGUE'' to refer 
to the two surveys generically or collectively.}
While SEGUE-1 included both
deep and shallow spectroscopic pointings, SEGUE-2 obtained only deep
pointings to better sample the outer halo, which is the primary
reason SEGUE-2 observed fewer stars than SEGUE.
Together, the SEGUE-1 and SEGUE-2 surveys comprise
358,000 stars observed along a grid of sightlines totaling 
2500 deg$^2$, with spectral
resolution $R \equiv \lambda/\Delta\lambda \approx 1800$
spanning $3800\mangstrom < \lambda < 9200\mangstrom$
(where $\Delta\lambda$ is the FWHM of the line-spread function).
Typical parameter measurement errors are $5-10\kms$ in radial velocity,
$100-200\,$K in $T_{\rm eff}$, and 0.21 dex in \feh,
depending on signal-to-noise ratio and stellar type 
(see \S\ref{sec:seg2}).
These data allow unique constraints on the stellar populations
and assembly history of the outer Galaxy and on the mass profile
of the Galaxy's dark matter halo.
SEGUE-2 observations are now complete.

SDSS-III also includes two bright-time surveys, generally performed when
the moon is above the horizon and the lunar phase 
is more than 70 degrees from new moon.
The first of these is the 
Multi-object APO Radial Velocity Exoplanet Large-area Survey
(MARVELS), which uses fiber-fed,
dispersed fixed-delay interferometer spectrographs
(\citealt{erskine00,ge02,ge02b,vaneyken10})  to monitor
stellar radial velocities and detect the periodic perturbations
caused by orbiting giant planets.
MARVELS aims to monitor 8,400 F, G, and K stars in the magnitude
range $V=8-12$, observing each star $\sim 24$ times over a 2--4 year
interval to a typical photon-limited velocity precision per observation
of $8\msec$ at $V=9$,
$17\msec$ at $V=10$, and $27\msec$ at $V=11$,
with the goal of achieving total errors within a factor 
of 1.3 of the photon noise.
These observations will provide a large and well characterized
statistical sample of giant planets in the period regime needed
to understand the mechanisms of orbital migration and
planet-planet scattering, as well as rare systems that
would escape detection in smaller surveys.
MARVELS began operations in Fall 2008 with a 60-fiber instrument,
which we hope to supplement with a second 60-fiber instrument
for the second half of the survey.  As of January 2011, it has obtained
more than 74,000 radial velocity measurements of 2580 stars.

The Apache Point Observatory Galactic Evolution Experiment
(APOGEE) will undertake an $H$-band ($1.51-1.70\micron$)
spectroscopic survey of $10^5$ evolved late-type stars
spanning the
Galactic disk, bulge, and halo, with a typical limiting (Vega-based)
magnitude of $H \approx 12.5$ per field.
Near-IR spectroscopy can be carried out
even in regions of high dust extinction, which will allow
APOGEE to survey uniform populations of giant/supergiant tracer stars
in all regions of the
Galaxy.  APOGEE spectra will have resolution $R \approx 30,000$,
roughly 15 times that of SEGUE-2,
and will
achieve a signal-to-noise ratio $\snr \ga 100$ per resolution element for
most stars.
These spectra will enable detailed chemical fingerprinting
of each individual program star, typically with 0.1-dex
measurement precision for $\sim15$ chemical elements that trace different
nucleosynthetic pathways and thus different populations of
progenitor stars.
Once APOGEE begins operations, MARVELS and APOGEE will usually
observe simultaneously, sharing the focal plane with fibers
directed to the two instruments, although this will not be
practical in all fields.
APOGEE will use a 300-fiber, cryogenic
spectrograph that is now (May 2011) being commissioned at APO.

SDSS-III will continue the SDSS tradition of releasing all data
to the astronomical community and the public, including calibrated
images and spectra and catalogs of objects with measured parameters,
accompanied by powerful database tools that allow efficient
exploration of the data and scientific analysis \citep{abazajian09}.
These public data releases will be numbered consecutively with those of
SDSS-I and II; the first is Data Release 8 (DR8; \citealt{dr8}), which
occurred in January 2011,
simultaneously with the submission of this paper.
To enable homogeneous analyses that span SDSS-I, II, 
and III, DR8 includes essentially all SDSS-I/II imaging and spectra,
processed with the latest data pipelines and calibrations.  DR8 
also includes all the new imaging data obtained for BOSS, and all
SEGUE-2 data.  
DR9, currently scheduled for Summer 2012, will 
include BOSS spectra obtained through July 2011,
new SEGUE stellar parameter determinations that incorporate
ongoing pipeline and calibration improvements,
and MARVELS radial velocity measurements
obtained through December 2010.
DR10, currently scheduled for July 2013,
will include BOSS and APOGEE spectra obtained through July 2012.
All data releases are cumulative.
The final data release, currently scheduled for December 2014,
will include all BOSS and APOGEE spectra and all MARVELS radial
velocity measurements.

The four subsequent sections describe the individual surveys in greater
detail.  We provide a short overview of the technical and scientific
organization of SDSS-III in \S\ref{sec:org}
and some brief concluding remarks in \S\ref{sec:summary}.

\section{BOSS}
\label{sec:boss}

According to general relativity (hereafter GR), the gravity of
dark matter, baryonic matter, and radiation should slow the expansion 
of the universe over time.  Astronomers attempting to measure this
deceleration using high-redshift Type Ia supernovae found instead
that cosmic expansion is accelerating
\citep{riess98,perlmutter99}, a startling discovery that had been
anticipated by indirect arguments
\citep[e.g.,][]{peebles84,efstathiou90,kofman93,krauss95,ostriker95,liddle96}
and has since been buttressed by more extensive supernova surveys and by
several independent lines of evidence
(see, e.g.,~\citealt{frieman08a} for a recent review).
Cosmic acceleration is widely viewed as one of the most profound
phenomenological puzzles in contemporary fundamental physics.
The two highest level questions in the field are:
\begin{enumerate}
\item Is cosmic acceleration caused by a breakdown of GR on
cosmological scales, or is it caused by a new energy component
with negative pressure (``dark energy'') within GR?
\item If the acceleration is caused by ``dark energy,'' is its
energy density constant in space and time and thus consistent
with quantum vacuum energy \citep{zeldovich68} or does its
energy density evolve in time and/or vary in space?
\end{enumerate}

For observational cosmology, the clearest path forward is to measure the
history of cosmic expansion and the growth of dark matter clustering over
a wide range of redshifts with the highest achievable precision,
searching for deviations from the model based on 
GR and a cosmological constant.
Supernova surveys measure the distance-redshift relation using
``standardized candles'' whose luminosities are calibrated by objects
in the local Hubble flow.  BOSS, on the other hand, employs a
``standard ruler,'' the BAO feature
imprinted on matter clustering by sound waves that propagate 
through the baryon-photon fluid in the pre-recombination universe
\citep{peebles70,sunyaev70,eisenstein98,meiksin99}.
The BAO scale can be computed, in absolute units, using straightforward
physics and cosmological parameters that are well constrained by cosmic
microwave background (CMB) measurements.
BAO are predicted to appear as a bump in the matter correlation function
at a comoving
scale corresponding to the sound horizon ($r=153.2\pm 1.7\,$Mpc,
\citealt{wmap7}),
or as a damped series of oscillations in the matter power spectrum
(see \citealt{eisenstein07b} for a comparison of the Fourier- and
configuration-space pictures).
When measured in the three-dimensional clustering of matter tracers at
redshift $z$, the transverse BAO scale constrains the angular diameter
distance $d_A(z)$, and the line-of-sight scale constrains the Hubble
parameter $H(z)$.

The first clear detections of BAO came in 2005 from analyses of
the 2dF Galaxy Redshift Survey \citep{cole05} and of the luminous
red galaxy sample (LRG; \citealt{eisenstein01}) of the SDSS
\citep{eisenstein05}.  The final SDSS-I/II BAO measurements determine
the distance to $z\approx 0.275$ with an uncertainty of 2.7\%
(\citealt{kazin10,percival10}; improved from the 5\% of \citealt{eisenstein05}).
Because of the leverage provided by this absolute distance
measurement, BAO measurements contribute substantially to the overall
cosmological constraints derived from SDSS galaxy clustering
(see \citealt{reid10}).

BOSS consists of two spectroscopic
surveys, executed simultaneously over an
area of 10,000 deg$^2$.  The first targets 1.5 million galaxies,
selected in color-magnitude space to be high-luminosity systems at large
distances.  The selection criteria, described further below, produce
a roughly constant comoving space density 
$n\simeq 3\times 10^{-4}\vunits$ to $z=0.6$, with a slight peak at
$z\simeq 0.55$, then a declining space density to $z\simeq 0.8$.
Relative to the SDSS-I/II LRG survey, which contained $10^5$ galaxies
out to $z=0.45$, 
the higher space density and higher limiting redshift of BOSS
yield an effective
volume (weighted by signal-to-noise ratio at the BAO scale) seven times
larger.\footnote{The SDSS main galaxy sample \citep{strauss02} contains
over 700,000 galaxies, but it has a median redshift of 0.1 and therefore
a much smaller effective volume for power spectrum measurements
on these scales.}  The second BOSS survey targets $1.5\times 10^5$ quasars,
selected from roughly $4\times 10^5$ targets (see below), in
the redshift range $2.2 \leq z \leq 4$, where \lya\ forest absorption
in the SDSS spectral range can be used as a tracer of high-redshift 
structure.\footnote{SDSS-I/II obtained spectra of 106,000 
quasars, but only 17,600 were at $z \geq 2.2$ \citep{schneider10}.}
The high density and large number of targets will allow BOSS to provide
the first ``three-dimensional'' measurements of large scale structure
in the \lya\ forest, on a sparsely sampled grid of sightlines that
collectively probe an enormous comoving volume.
The possibility of measuring BAO in the \lya\ forest was discussed
by \cite{white03}, and Fisher matrix forecasts were presented
by \cite{mcdonald07}, whose formalism was used to motivate and 
design the BOSS quasar survey.  While no previous survey has 
measured enough quasar spectra to reveal the BAO feature
in the \lya\ forest, analytic estimates and 
numerical simulations indicate that it
should be clearly detectable in the BOSS quasar survey
\citep{mcdonald07,slosar09,norman09,white10}.  
The characteristics of BOSS are summarized in 
Table~\ref{tbl:boss}.

Our forecasts, which are described in Appendix~\ref{appx:boss}, indicate that
BAO measurements with the BOSS galaxy survey should yield
determinations of $d_A(z)$ and $H(z)$ with $1\sigma$ precision of
1.0\% and 1.8\%, respectively, at $z=0.35$ (bin width $0.2<z<0.5$), 
and with precision of 1.0\%
and 1.7\%, respectively, at $z=0.6$ ($0.5 < z < 0.7$).  
The errors at the two redshifts
are essentially uncorrelated, while the errors on $d_A(z)$ and $H(z)$
at a given redshift are anti-correlated \citep{seo03}.
BAO are weakly affected by the effects of non-linear structure
formation, galaxy bias, and redshift-space distortions.
The primary consequence is a damping 
of oscillations in the power spectrum on small scales,
which can be well approximated by a Gaussian smoothing
\citep{bharadwaj96,crocce06,crocce08,eisenstein07b,
matsubara08a,matsubara08b,seo10,orban11}.
Our forecasts assume that density
field reconstruction \citep{eisenstein07a} can remove 50\% of the
non-linear Lagrangian displacement of mass elements
from their initial comoving locations \citep[e.g.][]{padmanabhan09a,noh09},
thereby 
sharpening the BAO feature and improving recovery of the original signal.
Forecasts with no 
reconstruction would be worse by factors of 1.6-2, while with perfect
reconstruction (not achievable in practice) they would improve by 
factors of $1.3-1.5$.
The uncertainty in BOSS BAO measurements is dominated by 
cosmic variance out to $z=0.6$; at these redshifts, 
a much higher density of targets (eliminating shot noise)
would decrease the errors by about
a factor of 1.4, while covering the remaining $3\pi$ steradians 
of the sky would reduce the errors by a factor of two.
Non-linear effects can also generate small shifts in the 
location of the BAO peak,
but current theoretical studies indicate that the statistical
errors will dominate systematic uncertainties associated with redshift space
distortions, non-linear evolution and galaxy bias
(see, e.g., \citealt{eisenstein07b,smith07,padmanabhan09b,
takahashi09,takahashi11}).
To allow some margin over our forecasts --- e.g., for reduced sky
coverage due to poor weather, or for problems in reconstruction,
or for other, unanticipated systematics ---
we have inflated our projected uncertainties by a factor of 1.2 
when defining the measurement goals reported in Table~\ref{tbl:boss}.

The \lya\ forest forecasts, performed with the
\cite{mcdonald07} formalism, indicate errors of 4.5\% and
2.6\%, respectively, on $d_A(z)$ and $H(z)$ at an effective
redshift $z \approx 2.5$ 
(with significant contributions from $2 \la z \la 3.5$).  
The errors are again anti-correlated:
the forecast error on an overall ``dilation factor''
that scales $d_A(z)$ and $H^{-1}(z)$ in proportion is only 1.8\%.
These predictions assume 15 quasars per deg$^2$ over
10,000 square degrees and no density field reconstruction. Reconstruction
is less important at high redshift and is unlikely to be possible
with a \lya\ forest survey as sparse as BOSS.
Our forecast calculations indicate that
the measurement precision is limited partly by the sparse sampling
of the density field and partly by the S/N of the spectra --- i.e.,
at fixed sky area, increasing either the exposure time per quasar
or the density of the quasar sample would decrease the errors.
However, given a fixed survey duration, the loss of sky area would
outweigh the gain from longer exposures, and the quasar surface
density is limited by our ability to efficiently select quasars
near the magnitude limit of SDSS imaging.

Our forecasts could prove somewhat optimistic, as broad 
absorption-line quasars may be unusable,
quasars observed in grey time will have lower signal-to-noise
spectra, and we have not included possible systematic uncertainties
associated with continuum determination, metal lines, or
damped \lya\ systems.  Conversely, use of additional imaging
data sets could improve quasar target selection in some areas
of the survey, increasing the surface density and improving
the BAO measurement precision.  Furthermore, these 
forecasts are based only on the location of the BAO peak as a 
function of angle with respect to the line of sight,
marginalizing away additional information contained in
the {\it amplitude} of \lya\ flux correlations as a
function of angle.  Including this information --- which
requires careful theoretical modeling to control systematics ---
could lead to significant (factor-of-two level) improvements
in the $d_A(z)$ and $H(z)$ constraints.  
More generally, the BOSS quasar survey
is pioneering a previously untried method of BAO measurement,
and performance forecasts are necessarily more uncertain than
for the galaxy survey.
\cite{slosar11} have used the first year of BOSS quasar
observations to make the first measurement of three-dimensional
large scale structure in the \lya\ forest.  While their 
measurements do not reach to the BAO scale, they detect flux
correlations out to at least $60\hmpc$ (comoving) and
find good agreement with predictions of a standard
$\Lambda$CDM cosmological model (inflationary cold dark matter
with a cosmological constant) out to this scale.

The underlying goal of these $d_A(z)$ and $H(z)$ measurements is
to probe the cause of cosmic acceleration, e.g., to
constrain the dark energy equation-of-state parameter $w$ and
its derivative $w_a$ with respect to expansion factor.
BOSS BAO measurements will also yield tight constraints
on other cosmological parameters, most notably the Hubble constant
$H_0$ and the curvature parameter 
$\Omega_k \equiv 1 - \Omega_m - \Omega_{\rm DE} - \Omega_{\rm rad}$.
Appendix~\ref{appx:boss} 
includes forecasts of BOSS constraints on these parameters
in combination with complementary data
(Table A1).  We also present forecasts incorporating the broad-band
galaxy power spectrum measurable with BOSS, which considerably
improves dark energy constraints. 
Controlling systematic effects on the broad-band power to 
extract the full statistical power of the data set will require
new work on the modeling of non-linear galaxy clustering
and bias.

\begin{figure*}
\epsscale{1.0}
\plotone{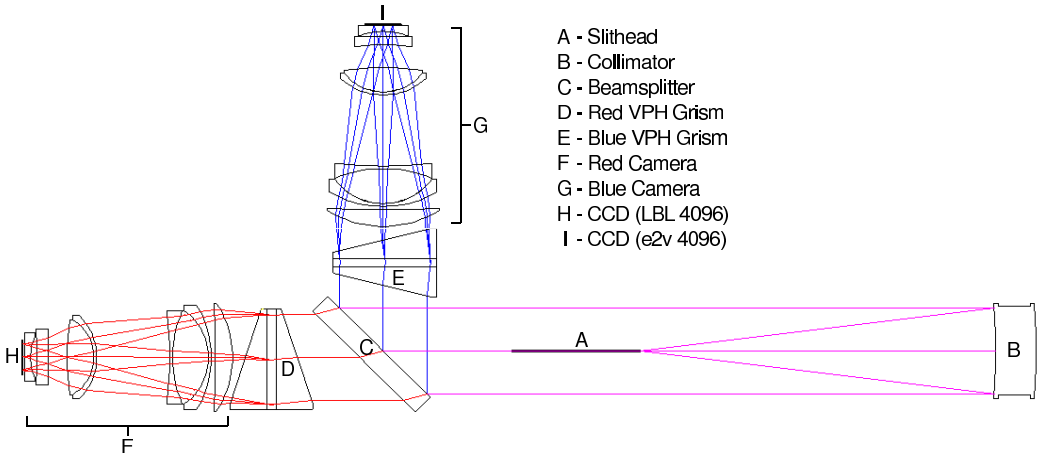}
\caption{
\label{fig:boss_spectrograph}
Schematic diagram of a BOSS spectrograph (one of two), with elements
as labeled.  The ``slithead'' is in fact a pseudo-slit containing
500 aligned fibers.
}
\end{figure*}

Since BOSS observes fainter targets than the original SDSS, it required
substantial upgrades to the two dual-channel spectrographs
\citep{york00}.
These upgrades were prepared during the first year of SDSS-III and
installed during the summer shutdown following completion of SEGUE-2.
In the red channel, the two $2048^2$, 24-micron pixel,
SITe CCDs were replaced
with $4128 \times 4114$, 15-micron pixel, fully-depleted,
250-micron thick devices from Lawrence Berkeley National Laboratory, with much
higher quantum efficiency at the reddest wavelengths, crucial for
galaxy redshift measurements at $z > 0.4$.
In the blue channel, the two $2048^2$ SITe CCDs were
replaced with $4096^2$, 15-micron pixel, e2v devices, with lower
read noise and greater sensitivity at the blue
wavelengths that are essential for \lya\ forest measurements. 
In both arms, the smaller pixel size and
larger format CCDs were selected to match the upgrade of the fiber
system from 640 fibers with $3''$ optical diameter
to 1000 fibers (500 per spectrograph) with $2''$ diameter.
The larger number of fibers alone improves survey efficiency by 50\%,
and because BOSS observes point sources (quasar targets) and distant
galaxies in the sky-dominated regime the smaller fibers yield
somewhat higher signal-to-noise spectra in typical APO seeing, 
though they place stiffer demands on guiding accuracy
and differential refraction.
The original diffraction gratings were replaced with higher throughput,
volume-phase holographic (VPH) transmission gratings from
Kaiser Optical Systems, and other optical elements were also
replaced or recoated to improve throughput.
The spectral resolution varies from
$\lambda/\Delta\lambda \sim 1300$ at $3600\mangstrom$ to 
3000 at $10,000\mangstrom$.
Figure~\ref{fig:boss_spectrograph} presents a schematic of one of the
BOSS spectrographs.
While we will not detail them here, we note
that the transition to BOSS also involved major upgrades to the
instrument and telescope control software, to the infrastructure
for fiber-cartridge handling, and to the guide camera, which was
replaced with an entirely new system.

\begin{figure}
\epsscale{1.0}
\plotone{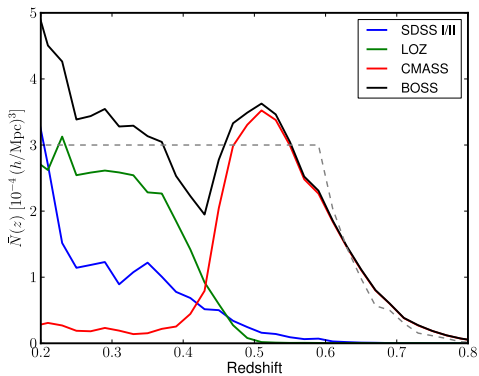}
\caption{
\label{fig:boss_galaxies}
The comoving space density of BOSS galaxies from data taken in Spring 2010.
The separate contributions of the LOZ cut, CMASS cut, and 
previously observed SDSS-I/II galaxies are shown, together
with the total.  The dashed curve shows our ``goal'' of constant
density to $z=0.6$ and tapering density beyond.  There is a
deficit near $z=0.45$ at the transition between the two
cuts, where obtaining accurate photometric redshifts for
target selection is difficult.
}
\end{figure}

BOSS galaxy targets are selected from the SDSS $ugriz$ imaging
\citep{fukugita96,stoughton02},
including the new imaging described below, using a series of
color-magnitude cuts.  These cuts are intended
to select a sample of luminous and massive galaxies
with an approximately uniform distribution of stellar masses
from $z\sim 0.2$ to $z\sim 0.6$. 
The sample is magnitude limited at $z>0.6$.  As in SDSS-I/II,
the selection is the union of two cuts designed to select
targets in two different
redshift intervals \citep{eisenstein01}.
Cut I (a.k.a.\ ``LOZ''), aimed at the interval $0.2 < z < 0.4$,
is defined by
\begin{equation}
r < 13.6  + c_{||}/0.3,\quad
| c_{\perp} | < 0.2, \quad
16 < r < 19.5.
\end{equation}
Cut II (a.k.a.\ ``CMASS'' for ``constant mass''), aimed at redshift $z > 0.4$,
is defined by
\begin{equation}
d_{\perp} > 0.55, \quad
i < 19.86 + 1.6 \times (d_{\perp} - 0.8), \quad
17.5 < i < 19.9 \,.
\end{equation}
The colors $c_{||}$, $c_{\perp}$ and $d_{\perp}$ are defined to track
a passively evolving stellar population with redshift,
\begin{eqnarray}
c_{||} &= 0.7 \times (g-r) + 1.2 \times (r-i-0.18) \\
c_{\perp} &= (r-i) - (g-r)/4 - 0.18 \\
d_{\perp} &= (r-i) - (g-r)/8~,
\end{eqnarray}
based on population synthesis models of luminous red galaxies
\citep{maraston09}.
The $r$-band and $i$-band magnitude limits are imposed using
{\tt cmodel} magnitudes \citep{abazajian04} rather than the $r$-band Petrosian
magnitudes used in SDSS-I/II \citep{petrosian76,strauss02}.  
(Both surveys used {\tt model} colors.)
The 215,000 galaxies observed by SDSS-I/II that pass these cuts
are included in the BOSS sample,
but they are not reobserved if they already had reliable redshifts.
Figure~\ref{fig:boss_galaxies}
shows the space density of BOSS galaxies (including the
SDSS-I/II objects) as a function of redshift, based on data
obtained through July 2010.
\cite{white11} have measured clustering in a sample of 
44,000 CMASS galaxies from the first six months of BOSS data
and used it to constrain the halo occupation distribution of
massive galaxies at $z=0.5$.  Their measurements confirm the
high clustering bias expected for such galaxies and assumed in our
BAO precision forecasts.

Because the BOSS BAO experiment uses quasars only as backlights for
the intervening \lya\ forest, there is no need to select the sample
homogeneously across the sky.  The quasar survey is allocated an
average of 40 targets per deg$^2$, and for \lya\ forest science the
essential criterion is to maximize the surface density of $z \geq 2.2$
quasars above the practical limit for BOSS spectroscopy ($g\approx
22)$.  Quasars at $z<2.2$ have little or no \lya\ forest in the
wavelength range covered by the BOSS spectrographs.
In detail, the ``value'' of a quasar for BAO studies is a
function of its redshift (which determines the observable \lya\ forest
path length) and its magnitude (which determines the S/N of the
spectrum).  Our recent studies on the SDSS southern equatorial stripe,
where deep co-added imaging and variability allow highly complete
identification of optically bright (``Type I'') quasars,
indicate that the surface density of $z \geq 2.2$
quasars to the BOSS magnitude limit is approximately 28 per deg$^2$
(see \citealt{palanque10}).
However, recovering these quasars from 40 targets per deg$^2$
in single-epoch SDSS imaging
is challenging because photometric errors 
are significant at this depth and because the quasar locus (in
$ugriz$) crosses the stellar locus at $z\approx 2.7$ \citep{fan99,
richards02}.  We therefore set the BOSS selection efficiency goal at 15
quasars per deg$^2$.  Any gains in selection efficiency above this
threshold translate into reduced errors on the BAO distance scale
measured from the \lya\ forest. 
Because the density field is sparsely sampled, the distance error
is (approximately) inversely proportional to the quasar surface density
at fixed survey area.

\begin{deluxetable}{l}
\tablecolumns{1}
\tablewidth{0pc}
\tablecaption{Summary of BOSS\label{tbl:boss}}
\tablehead{}
\startdata
{\it Duration:} Fall 2009 - Summer 2014, dark time\\
{\it Area:} 10,000 deg$^2$\\
{\it Spectra:} 1000 fibers per plate\\
\phantom{{\it Spectra:}} $3600\mangstrom < \lambda < 10,000\mangstrom$\\
\phantom{{\it Spectra:}} $R = \lambda/\Delta\lambda = 1300-3000$ \\
\phantom{{\it Spectra:}} $(\sn)^2$ \\
\phantom{{\it Spectra:}xxx} $\approx$ 22 per pix.\ at $i_{\rm fib}=21$ 
                            (averaged over 7000-8500\AA)\\
\phantom{{\it Spectra:}xxx} $\approx$ 10 per pix.\ at $g_{\rm fib}=22$ 
                            (averaged over 4000-5500\AA)\\
{\it Targets:} $1.5 \times 10^6$ massive galaxies, $z<0.7$, $i<19.9$ \\
\phantom{{\it Targets:}} $1.5\times 10^5$ quasars, $z \geq 2.2$, $g<22.0$ \\
\phantom{{\it Targets:}\ \ \ \ \ \ } selected from $4 \times 10^5$ candidates\\
\phantom{{\it Targets:}} 75,000 ancillary science targets, many categories \\
{\it Measurement Goals:} \\
\phantom{{\it Targets:}} Galaxies: $d_A(z)$ to 1.2\% at $z=0.35$ and 
                                     1.2\% at $z=0.6$\\
\phantom{{\it Targets:} Galaxies:} $H(z)$ to 2.2\% at $z=0.35$ and 
                                     2.0\% at $z=0.6$\\
\phantom{{\it Targets:}} \lya\ Forest: $d_A(z)$ to 4.5\% at $z=2.5$ \\
\phantom{{\it Targets:} \lya\ Forest:} $H(z)$ to 2.6\% at $z=2.5$ \\
\phantom{{\it Targets:} \lya\ Forest:} Dilation factor to 1.8\% at $z=2.5$\\
\enddata
\tablecomments{
BOSS imaging data were obtained in Fall 2008 and Fall 2009.
BOSS spectroscopy uses both dark and grey time (lunar phase $70-100$ degrees)
when the NGC is observable.
Galaxy target number includes 215,000 galaxies observed by 
SDSS-I/II.
Measurement goals for galaxies are 1.2 times the projected $1\sigma$
errors, allowing some margin over idealized forecasts.
Measurement goals for the \lya\ forest are equal to the $1\sigma$ forecast,
but this is necessarily more uncertain because of the novelty
of the technique.  The ``dilation factor'' is a common factor scaling
$d_A(z)$ and $H^{-1}(z)$ at $z=2.5$.
}
\end{deluxetable}

\begin{figure}
\epsscale{1.0}
\plotone{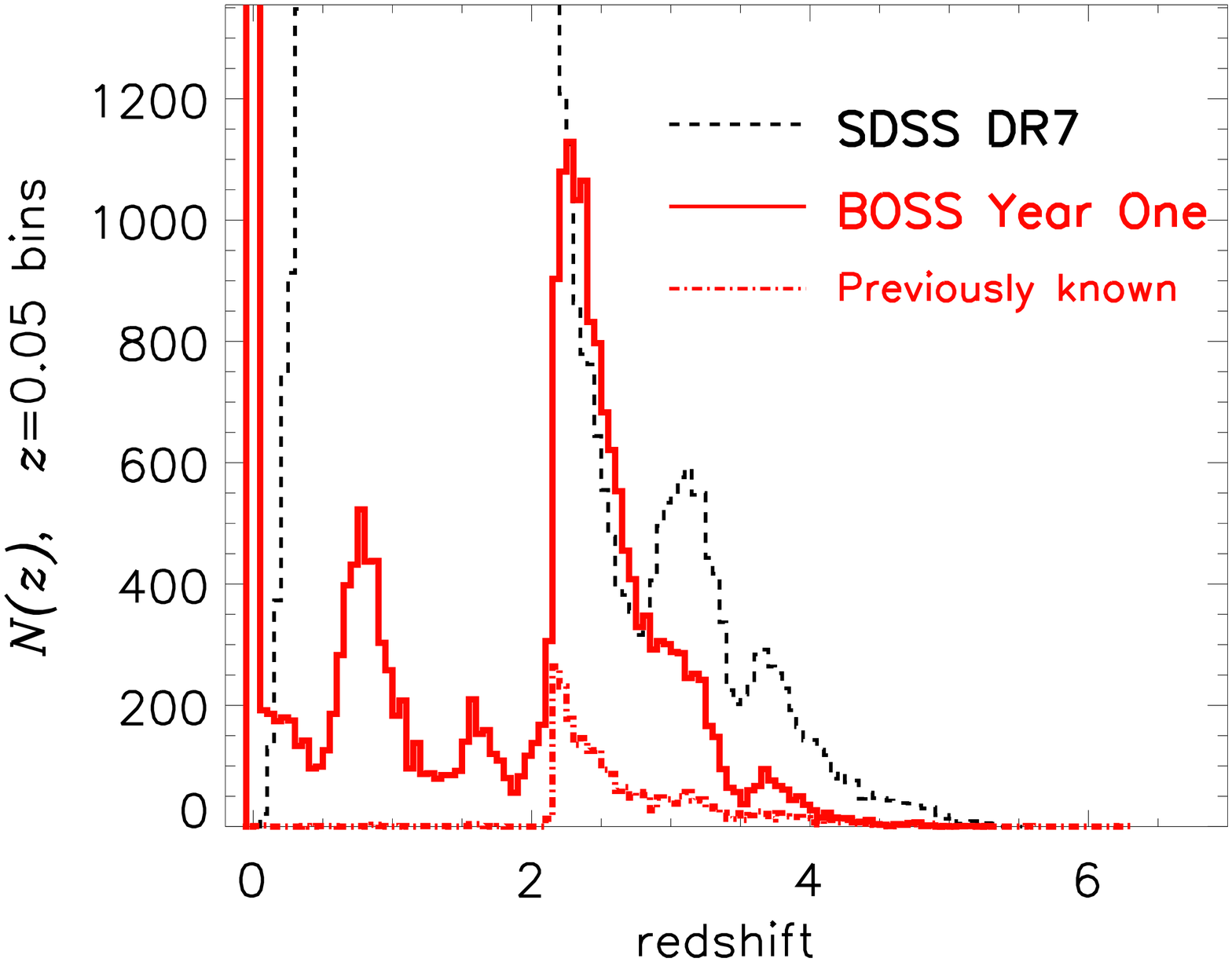}
\caption{
\label{fig:boss_qdist}
Redshift distribution of objects targeted by the BOSS quasar survey
and observed between December 2009 and July 2010 (red solid histogram).
There are 12,867 quasars with $z \geq 2.20$, obtained from a total
of 55,114 targets, of which 32,844 yielded reliable redshifts.
The spike at $z=0$ represents stellar contaminants, which are 34\% of
the objects with reliable redshifts.
For comparison, the black dotted histogram shows all quasars from the 
quasar catalog of SDSS DR7 \citep{schneider10}, and the red dot-dashed
histogram shows the previously known high-$z$ quasars in the
area surveyed, which come mostly but not entirely from DR7
and were reobserved by BOSS.
}
\end{figure}

Quasar science --- especially global population studies such as
luminosity functions, active black hole mass functions, and clustering ---
would 
benefit greatly from a homogeneous sample. We therefore select 20 of the 40
targets per deg$^2$ from single-epoch SDSS imaging using a ``core'' selection
method that remains fixed throughout the survey.  This core selection
is based on the probability, computed empirically from existing
survey data, that a given object is a high-redshift quasar rather than a 
star, low-redshift quasar, or galaxy 
\citep{bovy10,kirkpatrick11}.  The remaining
20 targets per deg$^2$, known as the ``bonus'' sample,
include previously known high-$z$ quasars (including those
from SDSS-I/II, reobserved to obtain higher S/N spectra),
FIRST radio sources \citep{becker95}
whose SDSS colors are consistent with $z \geq 2.2$, and objects
selected by a variety of methods including the KDE method of
\cite{richards09}, the neural network method of \cite{yeche10},
and lower priority likelihood targets.  These targets are selected
using additional data where they are available, including additional
SDSS epochs (which improve photometric precision where stripes overlap
and, on the southern equatorial stripe, provide variability information) and
photometry from GALEX (UV; \citealt{martin05}) and UKIDSS 
(near-IR; \citealt{lawrence07}).
The quasar selection criteria evolved significantly during
the first year of BOSS, as BOSS observations themselves provide vastly
more training data at these magnitudes than earlier surveys
such as 2SLAQ \citep{croom09} and AGES (C.\ Kochanek et al., in preparation).
The BOSS quasar target selection algorithms, including the
criteria used during the first year, are described in detail
by \cite{ross11} and the individual algorithm papers cited above.
With single-epoch SDSS imaging
we are presently achieving our goal of 15 quasars per deg$^2$,
improving to $\approx 18$ quasars per deg$^2$ where UKIDSS
and GALEX data are available \citep{ross11,bovy11}.
Figure~\ref{fig:boss_qdist} shows the
redshift distribution of BOSS quasars 
from spectra obtained
between December 2009 and July 2010; for this plot, all quasar
classifications and redshifts have been checked by visual inspection.
As of January 2011, BOSS has obtained spectra of 29,000 quasars
with $z \geq 2.2$ (according to pipeline redshifts), 
compared to $17,600$ from all of SDSS-I and II.

\begin{figure}
\epsscale{1.2}
\plotone{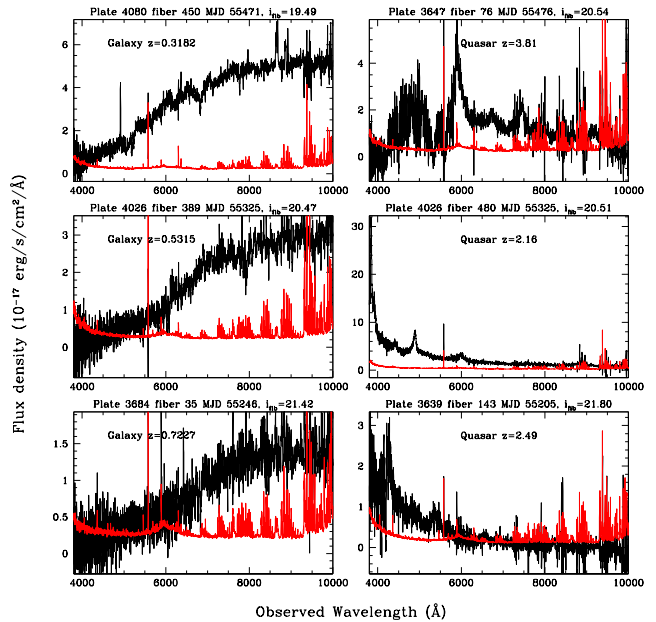}
\caption{
\label{fig:boss_spectra}
Examples of BOSS galaxy spectra (left) and
quasar spectra (right), smoothed with a 3-pixel boxcar.
In each panel, the black line is the spectrum and
the red is the estimated error per pixel.
The galaxy redshifts are
0.3182, 0.5315, and 0.7227 (top to bottom).  The calcium H\&K absorption
features are near 5200, 6200, and 6800 \AA\ (top to bottom).  
Other noticeable features are the Mgb absorption line and [OII]
and H$\alpha$ emission lines.
The quasar redshifts are 3.81, 2.16, and 2.49 (top to bottom).  The
\lya, CIV, and CIII] emission lines are identifiable features in these
quasar spectra.  The $2''$-fiber $i$-band magnitudes of the targets
are listed above each panel.
}
\end{figure}

Figure~\ref{fig:boss_spectra} shows several examples of BOSS galaxy
spectra (left) and quasar spectra (right), with brighter objects
at the top and targets near the magnitude limit at the bottom.
BOSS observations are done in a series of 15-minute exposures, with
additional exposures taken until a regression of $(\sn)^2$ against magnitude
(based on a fast reduction pipeline) yields 
$(\sn)^2 \geq 22$ per wavelength pixel ($1.4\mangstrom$)
at $i=21$ ($2''$ fiber magnitude) in the red cameras and
$(\sn)^2 \geq 10$ per wavelength pixel ($1.1\mangstrom$)
at $g=22$ in the blue cameras, where magnitudes are corrected for
Galactic extinction \citep{schlegel98}.\footnote{Higher 
$(\sn)^2$ thresholds, and consequently
longer exposure times, were employed during the first year.}
In transparent conditions, good seeing, and 
low Galactic extinction, the total exposure time is 45-60 minutes,
but the fixed $(\sn)^2$ criterion ensures homogeneity of redshift completeness
across the survey.  Our current data reductions, incorporating a 
spectroscopic reduction pipeline
adapted from the one originally developed for SDSS-I/II data by 
S.\ Burles and D.\ Schlegel, meets our science requirement of 95\%
redshift completeness for galaxy targets.  However, we plan to
implement the forward modeling
techniques described by \cite{bolton10} to extract
all the information contained in the spectra as accurately as
possible.  These pipeline improvements will increase our 
redshift completeness, improve galaxy science, and, most importantly,
yield higher S/N and better characterized errors in the \lya\ forest,
thus maximizing the return of the \lya\ forest survey.

\begin{figure}
\epsscale{1.2}
\plotone{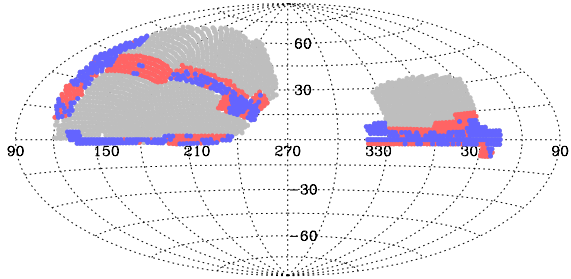}
\caption{
\label{fig:boss_footprint}
Planned footprint of the BOSS spectroscopic survey, showing
both the NGC (left) and SGC (right) regions.
Most of the imaging for SGC target selection was done as part
of SDSS-III.
Each circle marks the location of a spectroscopic plate.
Blue circles represent plates that have been observed as of
January 2011, while red circles represent plates that have
been drilled but not yet observed.  
The small extension of the SGC region below the equator 
at RA$>30^\circ$ is intended to reach the ``W1'' field
of the CFHT Legacy Survey.
}
\end{figure}

SDSS I and II imaged 7646 deg$^2$ of high-latitude sky in the
northern Galactic cap (NGC) and three stripes totaling 777 deg$^2$
of low extinction sky in the southern Galactic cap (SGC).\footnote{SDSS-II
also included 3200 deg$^2$ of lower latitude imaging for SEGUE,
but these data are not useful for BOSS.}  In order
to allow BOSS to cover 10,000 deg$^2$ with a balance between
the fall and spring observing seasons, BOSS used the SDSS camera
to image an additional 2500 deg$^2$ during the first 18 months
of SDSS-III, following the same procedures as SDSS I and II.
Figure~\ref{fig:boss_footprint} shows the full footprint for
BOSS spectroscopic observations.
The total area shown is 10,700 deg$^2$, while our science goal
for spectroscopy
is 10,000 deg$^2$; the exact breakdown between NGC and SGC 
in the spectroscopic survey will
depend on the amount of clear weather when these two regions
are observable.  Assuming historical weather patterns, we anticipate
a 5\% margin to complete the 10,000 deg$^2$ spectroscopic
survey by July 2014.

While our BAO measurement goals drive the design and the science
requirements of BOSS, the survey will enable a wide range of
other science.  
Redshift-space distortion analyses of BOSS galaxy clustering 
have the potential to yield strong constraints on clustering growth 
rates \citep{white09,reid11}, while weak lensing
by BOSS spectroscopic galaxies measured in SDSS (or deeper) imaging
can directly measure the evolution of matter clustering.
These methods could substantially increase the impact of BOSS in its
``core'' science area of testing theories of cosmic acceleration.
For large scale power spectrum measurements,
the much larger effective volume of BOSS (compared to SDSS-I/II)
will enable much stronger constraints on neutrino masses,
inflation parameters, and departures from ``vanilla'' $\Lambda$CDM.
BOSS galaxy spectra will provide a superb data set for studying
the evolution of massive galaxies from $z\approx 0.7$ to the present,
and they are expected to reveal $\sim 300$ examples of strong
gravitational lensing that can be used to constrain the mass profiles
of early type galaxies (e.g., 
\citealt{koopmans06,bolton08a,bolton08b,koopmans09}).
The high-redshift quasar data set will be ten times larger and approximately
2.5 magnitudes deeper (1.5 magnitudes at $z>3$) 
than SDSS-I/II, enabling much stronger
constraints on the evolution and clustering of quasars and the
black holes that power them.  The new BOSS imaging will extend
``tomographic''  studies of Milky Way structure
(e.g., \citealt{ivezic08}) and searches for ultra-faint dwarf galaxy
companions (e.g., \citealt{belokurov06}).

Finally, BOSS is devoting about four percent 
of its fibers to ``ancillary'' science
targets in a variety of categories.  These include studies of 
luminous blue galaxies at high redshifts,
brightest cluster galaxies,
star forming radio galaxies,
remarkable X-ray sources from {\it Chandra} and {\it XMM-Newton},
host galaxies of supernovae found in SDSS-II,
quasars selected by photometric variability,
double-lobed radio quasars,
candidate quasars at $z>5.6$,
variability in quasar absorption lines,
{\it Fermi} $\gamma$-ray sources,
distant halo red giants,
activity in late-M and L dwarfs,
hot white dwarfs,
and low mass binary star candidates.
Spectra from these ancillary science programs will be included
in the public data releases.

\section{SEGUE-2}
\label{sec:seg2}

The first SDSS imaging maps provided striking confirmation of
complex structure in the outer Galaxy \citep{ivezic00,yanny00,newberg02},
including the well known tidal tails of the Sagittarius dwarf galaxy
\citep{ibata94,ibata01,majewski03} and previously unrecognized streams, rings, 
and clumps (e.g., \citealt{odenkirchen01,
yanny03,grillmair06a,grillmair06b,juric08a}).
The ubiquity of this complex structure (e.g., \citealt{belokurov06,bell08})
supports the view that disrupted dwarf satellites are important 
contributors to the formation of
the Galactic halo \citep{searle78},
a scenario in qualitative and quantitative agreement with hierarchical,
CDM-based models of galaxy formation \citep{helmi99,bullock01,bullock05}.
These initial discoveries motivated the SEGUE-1 survey of SDSS-II
\citep{yanny09}, which included 3200 deg$^2$ of new $ugriz$ imaging
extending towards the Galactic plane and a spectroscopic survey
of 240,000 stars in a variety of target categories.
The first year of SDSS-III, during which the upgraded spectrograph
components for BOSS were being constructed, offered the opportunity
to roughly double the scope of SEGUE, using all of the dark time
over one year\footnote{Except for the time devoted to BOSS imaging.
SEGUE-2 also observed during grey time, with lunar phase
$70-100$ degrees from new moon.} 
instead of 1/3 of the dark time over three years.
In comparison to the Radial Velocity Experiment
(RAVE; \citealt{steinmetz06,zwitter08,fulbright10}), which 
has a roughly comparable number of stars,
SEGUE-1 and SEGUE-2 go much deeper
(to $g\sim 20$ vs. $I\sim 13$) and cover a larger 
wavelength range ($3800-9200\mangstrom$ vs.\ $8410-8795\mangstrom$),
but with lower resolution (1800 vs.\ 7500) and lower S/N. 
The SEGUE surveys probe the
distant disk and halo, while RAVE provides higher resolution
data concentrated in the solar neighborhood.

\begin{figure}
\epsscale{1.2}
\plotone{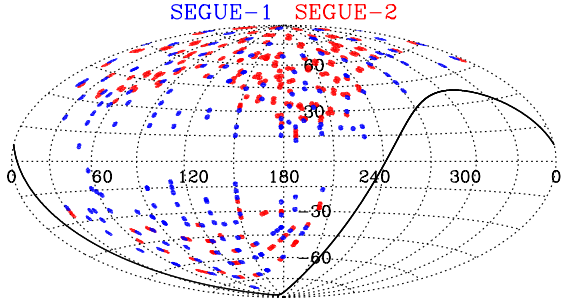}
\caption{
\label{fig:segue_fields}
Fields of the SEGUE-1 (blue) and SEGUE-2 (red) surveys, in Galactic coordinates.
The black curve marks $\delta = -20^\circ$.
}
\end{figure}

  The defining goal of SEGUE-2 is to map stellar populations and their
kinematics over a large volume of the Galaxy, from the inner halo out
to large Galactocentric distances
where late-time accretion events
are expected to dominate the halo population.  
SEGUE-1 and SEGUE-2
are similar enough in strategy and data quality to be treated
as a single data set.  Both surveys selected targets from the SDSS and SEGUE
$ugriz$ imaging data along individual 7 deg$^2$ lines of sight,
which are spread out over the imaging survey but do not form a
filled area.  Both surveys selected spectroscopic targets in several 
categories designed to map Galactic structure at different distances
or to identify classes of objects of particular astrophysical interest.
However, the target selection for SEGUE-2 is informed by the lessons 
from SEGUE-1.  The most
important strategic difference is that SEGUE-1 paired 
shorter exposures of brighter targets with deep 
spectroscopic pointings along 
the same sightlines, but SEGUE-2 obtained only deep pointings
so as to maximize coverage of the distant Galaxy.
The survey was designed to obtain 140,000 spectra, but worse than expected
weather led to a final sample of 118,151 stars.  
As with BOSS, SEGUE-2 exposures are accumulated until the S/N
crosses a pre-determined threshold.  For SEGUE-2, that threshold corresponds 
to median S/N$\,\approx 10$
per pixel ($\Delta\lambda \approx 1\mangstrom$)
for metal-poor main sequence turnoff stars at $r=19.5$ (PSF magnitude,
reddening corrected).
Under good conditions, reaching this S/N threshold required
approximately three hours of total exposure time.
Figure~\ref{fig:segue_fields} shows the distribution of SEGUE
and SEGUE-2 fields in Galactic coordinates.

  A detailed description of SEGUE-2 target selection will be provided
elsewhere (C.\ Rockosi et al., in prep.).  
The selection criteria for all the target categories
were adjusted based on what was learned from the SEGUE-1 data so as 
to obtain a higher success rate for categories like the low metallicity 
candidates and the blue horizontal branch
stars, or to push to larger mean distances for samples
like the halo main sequence turnoff stars.  In brief, the SEGUE-2 target 
categories,
selection criteria, and numbers of targets successfully observed are:
\begin{itemize}
\item Halo main sequence turnoff (MSTO) stars: $18 < g < 19.5$, 
      $+0.1 < g-r < +0.48$; 37,222 targets.
\item Blue horizontal branch (BHB) stars: $15.5 < g < 20.3$,
      $-0.5 < g-r < +0.1$, $+0.8 < u-g < +1.5$; 9983 targets.
\item K-giants: selected based on color and low proper motion,
      with $15.5 < g < 18.5$ and $r>15$; 42,973 
      targets.
\item M-giants: selected based on color and low proper motion, with
      $15.5 < g < 19.25$ and $i>14.5$; 631 targets.
\item Halo high velocity stars: selected based on color and high
      proper motion, $17 < g < 19.5$; 
      4133 targets.
\item Hypervelocity stars: selected based on color and high proper motion,
      $17 < g < 20$; 561 targets.
\item Cool extreme- and ultra-subdwarfs: selected based on color and
      reduced proper motion, with $15 < r < 20$ and $g>15.5$; 10,587 targets.
\item Low metallicity candidates: color selected, with $15.5 < g < 18$
      and $r>15$; 16,383 targets.
\end{itemize}
(Magnitude cuts are in PSF magnitudes.)
The first four categories are aimed primarily at understanding the
kinematic and chemical structure of the outer Galaxy, detecting
substructures in the stellar halo or outer disk, and constraining
the mass profile and shape of the Milky Way's dark matter halo.
These four categories have successively higher characteristic
luminosities, so they provide successively deeper but sparser probes,
with typical distance limits of 11 kpc (MSTO), 50 kpc (BHB) and
100 kpc (K/M-giants).
Hyper-velocity stars \citep{hills88,brown06} are thought to originate
in dynamical interactions with the Galaxy's central black hole,
and a systematic census of these stars can probe both the physics
of the ejection mechanism and the stellar
population at the Galactic Center.  \cite{kollmeier10}
present an analysis of this subset of the SEGUE-2 data.
The extreme subdwarf category is designed to find the most metal-poor
cool stars in the solar neighborhood.  
Finally, the low metallicity category aims to identify candidates
for future high-resolution spectroscopy that can probe
nucleosynthesis in the first generations of metal-poor or
metal-free stars.
Several target categories from SEGUE-1 were not extended to SEGUE-2, and their
fibers were redistributed to candidates of other spectral type.
The categories which have no targeted fibers in SEGUE-2 include
the white dwarf, ultra-cool white dwarf
\citep{gates04,harris08}, white-dwarf main sequence binary, and
G star categories.

\begin{deluxetable}{l}
\tablecolumns{1}
\tablewidth{0pc}
\tablecaption{Summary of SEGUE-2\label{tbl:segue2}}
\tablehead{}
\startdata
{\it Duration:} Fall 2008 - Summer 2009, dark+grey time\\
{\it Area:} 1317 deg$^2$, 118,151 targets\\
{\it Spectra:} 640 fibers per plate\\
\phantom{{\it Spectra:}} $3800\mangstrom < \lambda < 9200\mangstrom$\\
\phantom{{\it Spectra:}} $R = \lambda/\Delta\lambda = 1800$ \\
\phantom{{\it Spectra:}} $\sn \approx 10$ per pixel at $r_{\rm psf}=19.5$\\
{\it Target Categories:} \\
\phantom{{\it Targets:}} Halo main sequence turnoff stars (37,222)\\ 
\phantom{{\it Targets:}} Blue horizontal branch stars (9983)\\ 
\phantom{{\it Targets:}} K-giants and M-giants (43,604)\\ 
\phantom{{\it Targets:}} High-velocity stars (4133)\\
\phantom{{\it Targets:}} Hypervelocity stars (561)\\ 
\phantom{{\it Targets:}} Cool extreme subdwarfs (10,587)\\ 
\phantom{{\it Targets:}} Low metallicity candidates (16,383)\\ 
{\it Precision:} Dependent on stellar type and S/N, but typically\\
\phantom{\it Precision:} $150\K$ in $\teff$, 0.23 dex in $\log g$ \\
\phantom{\it Precision:} 0.21 dex in [Fe/H], 0.1 dex in [$\alpha$/Fe]\\
\enddata
\end{deluxetable}

\begin{figure}
\epsscale{1.2}
\plotone{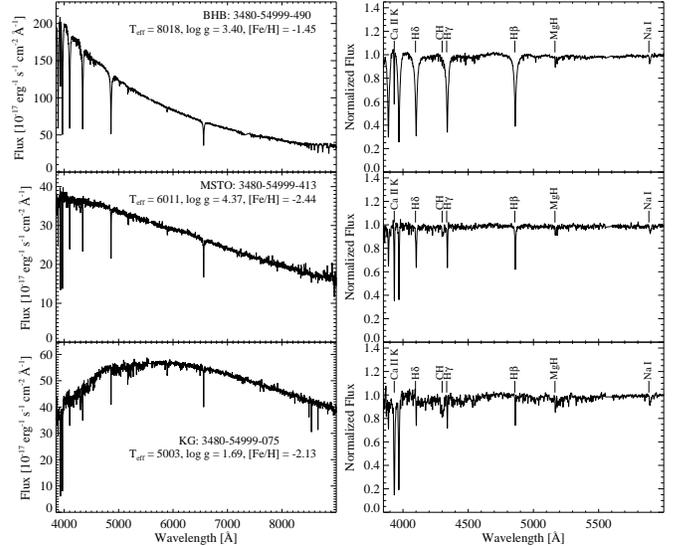}
\caption{
\label{fig:segue_spectra}
Example spectra 
of several classes of SEGUE-2 stars, all taken from the same
spectroscopic plate.  
The left-hand panels show the flux-calibrated spectra for a
halo blue horizontal-branch (BHB) star, a main-sequence turnoff (MSTO)
star, and
a K giant (KG) star, identified by 
plate-MJD-fiber. The stellar
atmospheric parameter estimates for each
object are shown below the label, as determined by the SSPP. The right-hand
panels
show the blue portion of each spectrum after continuum normalization
by the SSPP. Prominent spectral features in each spectrum are labeled.
Dereddened apparent magnitudes, from top to bottom, are
$g_0=16.27$, 17.82, 17.31 and 
$r_0=16.35$, 17.49, 16.67, respectively.
All of these stars are much brighter than the SEGUE magnitude limit.
}
\end{figure}

Figure~\ref{fig:segue_spectra} shows three
examples of SEGUE-2 stellar spectra: a BHB star (top),
a very metal-poor main-sequence
turnoff star (middle), and a very metal-poor K giant (bottom).
The left-hand panels show the flux-calibrated spectra over
the entire available spectral range, while the right-hand panels show the blue
portion of each spectrum after fitting and removing the continuum,
which aids examination of the detailed shape of the individual spectral
lines. The
estimated atmospheric parameters for each star, obtained as described
below, are displayed 
in the left-hand panels. Prominent spectral lines are identified in the
right-hand panels.

Like the SEGUE-1 spectra released in SDSS DR7,
SEGUE-2 spectra are first reduced by the 
{\tt idlspec2d} pipeline
(described in the DR8 paper, \citealt{dr8}), which performs sky subtraction
and wavelength and flux calibration, then extracts the one-dimensional 
spectrum,
carries out a basic classification, and measures the radial velocity.
This pipeline is unchanged from DR7.  The radial velocity
accuracy is $4\kms$ at $g$=18 (for detailed discussions, see
\citealt{yanny09} for SEGUE-1 and Rockosi et al., in prep., for SEGUE-2).
The SEGUE-2 stellar spectra are then processed
by the SEGUE Stellar Parameter Pipeline (SSPP; 
\citealt{lee08a,lee08b,allende08,smolinski10}),
and the three primary stellar
atmospheric parameters ($\teff$, log$\,g$, and [Fe/H]) are reported
for most
stars in the temperature range 4000 -- 10,000~K that have spectral S/N ratios
exceeding 10:1 per \AA\ 
(averaged over the entire spectrum). The SSPP estimates stellar
atmospheric parameters using several approaches, such as a
minimum distance method 
\citep{allende06}, neural network analysis
\citep{bailer-jones00,willemsen05,refiorentin07},
auto-correlation analysis 
\citep{beers99}, and a variety of line index
calculations based on previous calibrations with respect to known standard
stars
\citep{beers99,cenarro01a,cenarro01b,morrison03}.
We refer the interested reader to 
\cite{lee08a} for more details on the SSPP and to
\cite{smolinski10} and \cite{dr8} for a description of 
recent updates. The current best estimates of the precision of the
derived parameters $\teff$, log $g$, and [Fe/H] are 150~K, 0.23 dex, and
0.21 dex, respectively, for SEGUE
stars with 4500~K $\leq T_{\rm eff} \leq$ 7500~K.
As described by \cite{lee10},
the SSPP has recently been extended to provide estimates of
alpha-element-to-iron
ratios, with precision of $\sim 0.1$ dex in [$\alpha$/Fe] 
for stars with $\sn > 20:1$,
4500 $\leq T_{\rm eff} \leq$ 7000~K, and 
[Fe/H] $> -2.5$ (or as low as $-3.0$ for cooler stars). 
\cite{lee11} use these measurements to chemically separate ``thin disk''
and ``thick disk'' populations and thereby measure their kinematics
with unprecedented precision.
Validation and refinements of the SSPP parameter estimates are still underway,
based on a uniform re-analysis of more than one
hundred high-resolution spectra of
SDSS/SEGUE stars obtained with the Hobby-Eberly, Keck, Subaru, and VLT
telescopes. 

\section{APOGEE}
\label{sec:apogee}


No previous spectroscopic survey of 
the chemistry of Milky Way stars has included
all Galactic stellar populations in a systematic, consistent manner.
This limitation arises
primarily because the inner regions of the Galaxy are largely
hidden at optical
wavelengths by obscuring interstellar dust.  Furthermore, obtaining
precise, element-by-element abundances requires high-resolution,
high-S/N spectra, well beyond the capabilities of SEGUE or of
any large spectroscopic survey to date.  APOGEE will address
both of these limitations by obtaining $H$-band ($1.51-1.70\micron$) spectra
for $10^5$ late type, evolved stars with
a FWHM resolving power $R = \lambda/\Delta\lambda \approx 30,000$ 
and a minimum S/N of $100$ per resolution element.
Thanks to the greatly reduced extinction at
infrared wavelengths ($A_H/A_V = 1/6$), APOGEE will
observe efficiently even in heavily obscured regions of the Galaxy.  It will
be the first large scale, systematic, high-precision
spectroscopic survey of {\it all} Galactic stellar populations ---
those in the bulge, bar, thin disk, thick disk, and halo ---
conducted with a homogeneous
set of spectral diagnostics, data quality and stellar tracers.

Tests on simulated spectra indicate
that APOGEE will deliver radial velocities with an accuracy
of $0.3\kms$ or better and star-by-star abundances of $\sim$15
key elements, with $\sim$0.1-dex precision (for solar metallicity targets),
including the most common metals C, N and O, many of the
$\alpha$-elements, several iron-peak elements, and two odd-$Z$ elements,
Na and Al.
These different species form through different nucleosynthetic pathways
in stars
of different mass and metallicity, and they therefore provide complementary
information about chemical evolution of their parent galaxy
(the Milky Way or a dwarf progenitor), about
the physics of stellar
and supernova nucleosynthesis, and about the mixing and enrichment
history of the interstellar
medium in the Galaxy.  
APOGEE will increase the total number of
high resolution, high-S/N stellar spectra obtained under
uniform conditions at any
wavelength by more than two orders of magnitude.


APOGEE's main science objectives are:
\begin{itemize}
\item To derive tight constraints on models for the history of star
formation, chemical evolution (including the processes of
chemical mixing and feedback in the interstellar medium
and dredge-up in individual stars), and mass assembly of the Galaxy. 
\item To constrain the stellar initial mass function (IMF) in each of the main
Galactic components.
\item To derive kinematical data at high precision useful for
constraining dynamical models of the disk, the bulge, the bar,
and the halo and for discriminating substructures in these
components, if/where they exist.
\item To infer properties of the earliest stars (usually
thought to reside or to have resided in the Galactic bulge), by detecting
them
directly if they survive to the present day in significant numbers
or by measuring their
nucleosynthetic products in the most metal-poor stars that do survive.
\item To unravel the overall formation mechanisms and evolution of the 
Milky Way by coupling the extensive chemical data to the dynamics of the
stars.
\end{itemize}

The APOGEE sample size will be large enough to provide statistically
reliable measures of chemistry and kinematics in dozens of separate
zones defined by cuts in Galactocentric radius, Galactic longitude,
and height above the plane,
at the level of precision currently available only for stars in the solar
neighborhood.  Detailed chemical fingerprints will allow identification
of sub-populations that have a common origin but may now be
distributed widely around the Galaxy, providing unique insights into
the Galaxy's dynamical history.
In more general terms, APOGEE and SEGUE provide the kinds of data
needed to use the Milky Way as a detailed test of contemporary
galaxy formation theory (see, e.g., \citealt{freeman02}).

\begin{figure*}
\epsscale{1.0}
\plotone{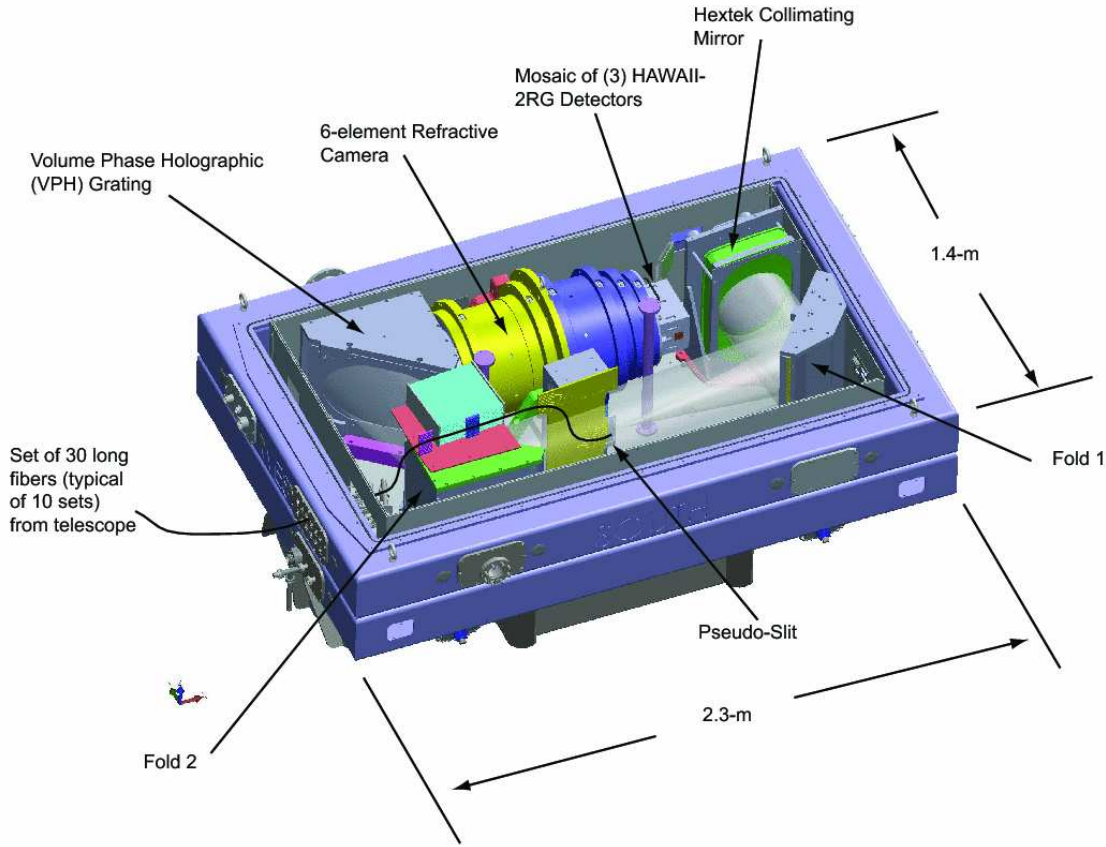}
\caption{
\label{fig:apogee_schematic}
A SolidWorks(TM) model rendering of the
fiber-fed APOGEE spectrograph.  300 fibers are arranged in a column on a
pseudo-slit.  Divergent light from the fiber ends is collimated and
redirected back towards the pseudo-slit by a collimating mirror.  The
collimated light passes through the pseudo-slit and is dispersed by a volume
phase holographic grating and refocused by a six-element camera onto a mosaic
of three HAWAII-2RG detectors.  Fold mirrors are used to package the
optical train in an efficient manner.
}
\end{figure*}

The fiber plugplate system for APOGEE is similar
to that used for BOSS and SEGUE-2, but the APOGEE spectrograph
sits in a building adjacent to the Sloan telescope rather than
being mounted on the telescope underneath the mirror.  
For APOGEE, 300 ``short'' (2.5-meter)
fibers
carry light from the focal plane to a connection below the telescope,
where they are coupled to ``long'' (40-meter) fibers that
transport the signal to the spectrograph enclosure, penetrate the
evacuated dewar, and illuminate a pseudo-slit.
The fibers have an outer diameter of $190\micron$ and an inner
(light-transmitting)
diameter of $120\micron$, corresponding to $2''$ on the sky.
Figure~\ref{fig:apogee_schematic} presents a SolidWorks (TM)
model rendering of the APOGEE spectrograph \citep{wilson10}.
The key elements are:
\begin{itemize}
\item A 2.1 m$^3$ cryostat, maintained at approximately $80\,$K
and a vacuum level below $10^{-6}\,$Torr.
\item A $50.8\cm\,\times\,30.5\cm$ volume phase holographic (VPH) grating, 
the largest yet employed in an astronomical instrument,
manufactured with a novel mosaic process by Kaiser Optical Systems.
\item A six-element camera manufactured and aligned by New England
Optical Systems, with optical elements of silicon and fused silica
to a maximum diameter of 40 cm.
\item Three 2K$\times$2K HAWAII-2RG infrared detectors
\citep{garnett04}
mounted on a translational stage that enables 
precise subpixel dithering in the dispersion dimension.
\item A collimator with electromechanical control for focusing and dithering
in the spatial dimension (useful for making high quality, ``smeared"
flatfields).
\item Two fold optics used to create a compact spectrograph design and,
through the use of dichroic coating on Fold Mirror 2, 
to divert and sequester out-of-band light.
\end{itemize}
These components were integrated, aligned, and tested at the
University of Virginia. The full spectrograph is now (May 2011) 
being commissioned at APO.

The three detectors span the wavelength range $1.51-1.70\micron$ with
two small gaps.  The optical design spectral resolution increases from
$R \equiv \lambda/\Delta\lambda \approx 27,000$ at the
shortest wavelengths to $R\approx 31,000$ at the longest wavelengths.
The pixel scale is $0.35$-$0.23$\AA\  (blue to red spectral end),
so there are typically 1.6 pixels per FWHM 
at the blue end and 2.3 pixels per FWHM at the red end.
APOGEE observations will consist of pairs of exposures
dithered by half a pixel, which will then be combined to yield
fully sampled spectra at the instrument resolution at all wavelengths.
In a typical observation, $\sim$250 fibers
will be devoted to science targets and $\sim$50 to calibration stars,
telluric standards, and sky.
The throughput requirement on the spectrograph is to achieve
$\snr=100$ per resolution element in three hours of exposure time under
good conditions for a star with Vega magnitude $H=12.2$.
Our measurements of component throughputs and our early commissioning
data both suggest that the total
throughput will be somewhat better than this level and will
reach the above S/N for $H=12.5$ stars in three hours.
Visits to each star field will be about one hour, so that
most program stars will be observed at least three times.
Because internal binary
velocities can distort measurements of Galactic kinematics (e.g., by
inflating velocity dispersions), the cadences of these repeat visits will 
be designed so that the majority of the
most troublesome binaries can be identified via radial velocity variations.

\begin{deluxetable}{l}
\tablecolumns{1}
\tablewidth{0pc}
\tablecaption{Summary of APOGEE\label{tbl:apogee}}
\tablehead{}
\startdata
{\it Duration:} Spring 2011 - Summer 2014, bright time\\
{\it Area:} $\sim 1575$ deg$^2$ \\
\phantom{{\it Area:}} $\sim$ 230 fields \\
\phantom{{\it Area:}} total exposure times 1-24 hours\\
{\it Spectra:} 300 fibers per plate\\
\phantom{{\it Spectra:}} $1.51\micron < \lambda < 1.70\micron$ \\
\phantom{{\it Spectra:}} $R \equiv \lambda/\Delta\lambda = 27,000 - 31,000$\\
\phantom{{\it Spectra:}} typical $\sn \geq 100$ per resolution element\\
{\it Targets:} $10^5$ 2MASS-selected, evolved
stars with $(J-K_s)_0 > 0.5$\\
\phantom{{\it Targets:}} Includes RGB, clump giants, AGB, red supergiants\\
{\it Precision:} Abundances to 0.1 dex internal precision \\
\phantom{{\it Precision:xxx}} 0.2 dex external precision\\
\phantom{{\it Precision:}} $\sim 15$ elements per star \\
\phantom{{\it Precision:}} Radial velocities to $<0.3$ km s$^{-1}$
per visit.\\
\enddata
\end{deluxetable}


APOGEE and MARVELS share the focal plane during bright time observations,
with separate fibers on the same 7 deg$^2$
plugplates feeding the two instruments.
This scheme nearly doubles the observing time available to both surveys,
at the cost of reducing flexibility.  In particular, MARVELS
observations require visiting the same field as many as $24$ times
throughout the 3-year APOGEE survey.
APOGEE will therefore devote most of its
observations to ``long'' fields, observed for a total of 24 hours
or, in cases where earlier MARVELS observations have already
accumulated many epochs of data, for smaller total visit times (e.g.,
10 hours).  The nominal APOGEE exposure time is three hours, but
most fields have many more than 250 potential APOGEE targets, 
so the additional observing time can be used to increase the 
number of stars in a given field
by up to a factor of eight.
It can also be used 
to increase depth by observing fainter stars for total exposure
times as long as 24 hours, and perhaps to obtain higher \snr\ spectra
for lower metallicity stars that have weaker lines.
The detailed mix among these strategies is not yet decided.
Roughly 25\% of the observing time will be assigned to APOGEE-only,
``short'' fields
that increase the sky coverage of the survey, with three one-hour
visits in disk and halo fields and single one-hour visits (concentrating
on brighter stars) in bulge
fields that are only available for short periods at acceptable airmass from
APO.

\begin{figure}
\epsscale{1.2}
\plotone{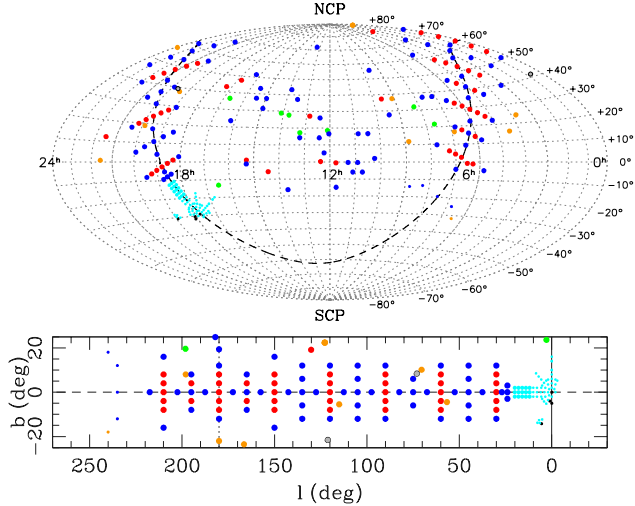}
\caption{
\label{fig:apogee_fields}
Current plan of the APOGEE field distribution
in equatorial (top) and Galactic (bottom) coordinates. Fields
are represented with circles scaled to represent their actual size
(typicaly 7 deg$^2$). Red circles denote ``24-hr'' 
APOGEE/MARVELS fields.
Green circles denote ``10-hr'' APOGEE/MARVELS fields. 
Orange circles denote fields
containing calibration clusters, which will typically be observed for a
total of three hours. Two special 6-hr fields are shown in grey. APOGEE-only 
fields (also observed for only three hours) are shown with blue circles. 
Fields targeting
the inner disk and bulge and the core of the Sagittarius dSph galaxy
will have a smaller field of view ($\sim 3$ or $\sim 1$ deg$^2$) 
to compensate for being
observed at high airmass (see text) and are shown as either 3-hr
(black) or 1-hr (cyan) fields. The heavy dashed line in the upper plot
marks the Galactic plane.
}
\end{figure}

Figure~\ref{fig:apogee_fields} shows the currently planned
distribution of APOGEE fields on the sky, although the plan
remains subject to (likely minor) changes.  Some fields target globular or
open star clusters, both for science investigations and to
allow calibration of APOGEE abundances against previous data sets 
for these clusters.
Most fields are designed to provide systematic coverage of
the Galaxy, subject to the constraints of observability.
For all fields, target selection is based primarily on
2MASS photometry \citep{skrutskie06}, 
with data from {\it Spitzer}/IRAC \citep{churchwell09}
and {\it WISE} \citep{wright10}
used to correct for dust reddening on a star-by-star basis
(see \citealt{majewski11}).
At low latitudes, APOGEE targets will be selected
with a simple, dereddened, near-infrared color limit
---  e.g., $(J-K_s)_0 > +0.5$; such a sample will be dominated
by late type, evolved stars like red giant branch, asymptotic giant
branch, red clump, and red supergiant stars.  These stellar species
span all population ages and allow APOGEE
to probe to the farthest distances at a given magnitude.
At higher latitudes where the ratio of distant giants to nearby dwarf stars
is lower, the simple $(J-K_s)_0$
color selection will be supplemented by newly obtained photometry in the
Washington $M$, $T_2$ and $DDO51$ filters from the
U.S. Naval Observatory 1.3-m telescope, which will greatly
reduce the contamination of the APOGEE target
sample by nearby dwarf stars (see \citealt{geisler84,majewski00}).
Because of the large differential refraction 
across the nominal 3$^{\circ}$ field of view when observing at large airmass,
APOGEE fields targeting the inner disk and bulge and the core of the
Sagittarius dSph galaxy will be observed with fibers placed in
a reduced (e.g., 1$^{\circ}$) field of view.  Because of the high
stellar density in these fields, there will be no shortage of targets
to fill the entire complement of fibers.

In each field, targets will be selected in three different
bins of $H$-band magnitude to provide a reasonable spread
in distance coverage.  In the
absence of dust extinction, a typical red clump giant has
$H=12.5$ at a distance of $\sim$6 kpc, while a typical
star near the tip of the red giant branch has $H=12.5$ at a
distance of 40 kpc.  Observations to $H=13.5$ (achievable
for some targets in long fields) reach a factor 1.6
farther in distance.  While extinction in the $H$-band is much
lower than in the optical, dust can still 
significantly reduce the survey depth in the inner disk and bulge
in the direction of dense clouds.
The dust distribution in the Galactic mid-plane is patchy on the
3$^{\circ}$ scale of the 2.5-m field-of-view, which increases the incidence
of lower reddening windows in the targeted fields.  
However, even
where $A_V=10$, APOGEE will be able to probe to the far 
edge of the Galactic disk with RGB tip stars in ten hours of integration.

Deriving chemical abundances from $10^5$ high-resolution spectra
presents a major analysis challenge.	We are developing an optimal
spectral extraction and calibration pipeline and a stellar
parameters and chemical abundance pipeline, and we have created realistic
simulated data to test these pipelines in advance of APOGEE
observations.
The extraction pipeline performs a number of tasks, including
bundling of hundreds of up-the-ramp detector reads for each
pixel,\footnote{The
HAWAII-2RG detectors in the APOGEE instrument can be operated with
non-destructive readouts.  
APOGEE will read the collected charge in each pixel
at regular, frequent intervals (``up-the-ramp sampling") throughout the
integration and use this to reduce readout noise and to monitor
pixels for the incidence of cosmic rays, the onset of pixel-well charge
saturation,
and the progress of each integration in the presence of variable observing
conditions.}
using the up-the-ramp detector operation to correct for cosmic rays and pixel
saturation, and performing
sky subtraction, two-dimensional to one-dimensional spectral
extraction, wavelength
calibration, combination of dithered exposure pairs into single, fully
sampled
spectra, telluric absorption correction, flux calibration,
and the measurement of stellar radial velocities.
Spectra from multiple visits are individually corrected to
rest-frame wavelengths and optimally combined for each star.

Parameter and abundance determination then proceeds in two stages.
First, spectral fitting based on $\chi^2$ minimization is used to
constrain the stellar temperature
($T_{\rm eff}$), the surface gravity ($\log g$), the microturbulence parameter
($\xi$),
and the abundances of elements that have an important effect on
stellar atmospheric structure ---
$[{\rm Fe/H}]$,
$[{\rm C/H}]$, and
$[{\rm O/H}]$.
Other abundances are then determined by one-at-a-time $\chi^2$
minimization with the former parameters held fixed.
We plan on using
Kurucz \citep{castelli04}, MARCS \citep{gustafsson08}, and, for the warmest
targets,  TLUSTY \citep{lanz07} model atmospheres.
A variety of literature sources are being used
to create and cross-check $H$-band line lists, using theoretical
or laboratory $gf$-values when they are available, and otherwise
inferring semi-empirical ``astrophysical" $gf$-values from fitting
synthetic spectra to
high-resolution observations of standard stars, such as the Sun and Arcturus.

\begin{figure}
\epsscale{1.1}
\plotone{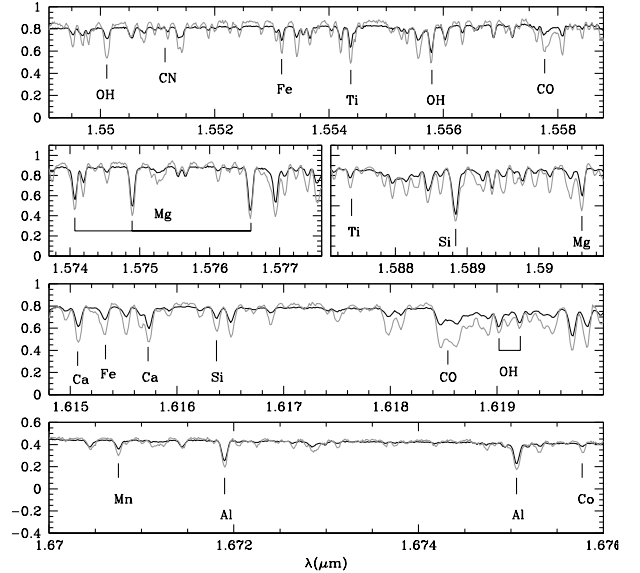}
\caption{
\label{fig:apogee_spectrum}
Selected sections of
simulated APOGEE spectra for two giant stars with $T_{\rm eff}$=4000~K, $\log
g$=1.0 and [Fe/H]=$-1.0$ (solid line) and 0.0 (grey line) showing some of
the absorption 
lines that will be used to measure element abundances. The vertical
scale is in $F_\lambda$ units 
(erg ${\rm cm^{-2}}$ ${\rm s^{-1}}$ ${\rm\AA^{-1}}$), 
with an arbitrary normalization.
The plotted regions cover only $\sim 15\%$ of the full APOGEE spectral range.
}
\end{figure}

Figure~\ref{fig:apogee_spectrum} shows simulated spectra at
APOGEE resolution, sampling, and \snr\
for two giant stars with $T_{\rm eff}$=4000~K, $\log
g$=1.0 and [Fe/H]$=-1.0$ (solid line) and 0.0 (grey line).  The spectral
regions displayed (totaling only $\sim15$\% of the total APOGEE spectral
range)
sample some of the absorption lines that will be used to determine
elemental abundances from APOGEE spectra, including Fe, the key metals
C, N, and O (which will be
determined from OH, CO, and CN lines), Al, Mn, Co, and several $\alpha$
elements (Mg, Si, Ca, Ti).

\section{MARVELS}
\label{sec:marvels}

Over the last 15 years, the study of extra-solar planets has advanced
from first discoveries \citep{wolszczan92,mayor95,marcy96} to large
surveys that have revealed an astonishing diversity of planetary
systems. The standard, core accretion scenario of giant planet
formation (see, e.g., \citealt{lissauer87}) predicts that planets like
Jupiter form in nearly circular orbits in the region beyond the ``snow
line'' in the protoplanetary disk where ices are stable, corresponding
to orbital periods of several years or more for solar-type stars.  The two
greatest surprises of extra-solar planetary discoveries to date have
been that many giant planets have periods below one year, sometimes as
short as one day, and that many of these planets are on highly
eccentric rather than circular orbits.	The first finding suggests
that many giant planets ``migrate'' inward after their formation,
while the second suggests that some dynamical mechanism must excite
the planetary eccentricities, probably after the protoplanetary disk has
dispersed. Various mechanisms have been proposed to explain planetary
migration and the broad eccentricity distribution.  
These include ``smooth'' migration via interaction with the 
proto-planetary gas or planetesimal disk \citep{lin96,murray98}, 
violent migration via
dynamical processes such as planet-planet scattering
\citep{rasio96,weiden96,ford08,juric08},
eccentricity pumping via the Kozai mechanism
\citep{holman97,wu03,fabrycky07}, and
tidal circularization of
the highest eccentricity systems to explain the shortest-period giant
planets \citep{wu03,fabrycky07,nagasawa08}.  It is unclear which, if any,
of these mechanisms dominate, and what fraction of systems escape migration
and remain on nearly circular orbits.

A large and well characterized sample of giant planets with
periods less than a few years is essential for solving the riddles of
migration and orbital eccentricities.
When coupled with detailed {\it ab initio} simulations of planet formation,
the various proposed migration mechanisms make different predictions for the
resulting (post-migration) distributions of planet masses, semimajor
axes, and eccentricities.
Comparison to the observed
distribution of these properties thereby constrains
the physical processes involved in planet formation and migration.
With the largest homogeneous and statistically complete sample
of planets that is currently available \citep{cumming08}, it is 
possible to place constraints on a few specific or extreme migration scenarios
(e.g., \citealt{schlauf09}), but a substantially larger sample is needed
to draw strong conclusions \citep{armitage07}.

The new generation of planet search experiments are using a variety
of technical approaches --- high-precision radial velocities (RV),
transits, microlensing, and direct imaging --- 
to push forward along several distinct
dimensions of parameter space, including lower masses and longer
periods.  Uniquely among these experiments, MARVELS focuses on 
greatly expanding the target sample for giant planets
(roughly speaking, Jupiter mass and larger) in the
short-to-intermediate period regime that is most critical
for understanding migration and dynamical interaction.
It exploits the novel capabilities of fixed-delay dispersed
interferometers, which combine interferometers with moderate-resolution
spectrographs to enable precision RV measurements with high
throughput and a relatively small amount of detector real estate 
\citep{erskine00,ge02,ge02b}.
This method enables highly multiplexed, multi-fiber searches on moderate
aperture telescopes \citep{ge02,ge03}, allowing MARVELS to
move much further than previous experiments towards 
large and relatively unbiased target samples.

The basic principles of RV measurement with a dispersed
fixed-delay interferometer (DFDI) are reviewed by 
\cite{ge02} and \cite{vaneyken10}.  In brief,
light from the telescope (either a stellar source or a velocity-stable
calibration source) is first fiber-fed to a Michelson interferometer. One of
the interferometer mirrors is angled such that the optical path difference
(in units of waves) changes as a function of vertical height above the 
optics bench, in addition to depending on wavelength.
Putting this interferometer output through a slit and a spectrograph
produces, at the two-dimensional detector, an intensity pattern of
constructive and destructive interference that appears as diagonal lines
(the interferometer ``comb''). For an absorption line source, this comb
is multiplied into the absorption lines, creating a moir\'e pattern of
intersections between the diagonals and the vertical absorption lines. If
the diagonal lines are close to vertical (high slope), then a small shift
in absorption line wavelength due to radial velocity change is multiplied
by the slope to create a large shift in the vertical intersections between
the comb and the absorption lines. Thus, even if the spectrograph resolution is
too low to permit accurate measurements of the horizontal line shifts in
wavelength, the amplified vertical fringe shifts can be measured accurately.
In technical terms, the combination of an interferometer with a
spectrograph heterodynes high frequency spectral information to lower
frequencies that survive blurring by the moderate resolution 
spectrograph without losing the Doppler signal needed for 
precision RV measurement 
(see \citealt{wang11}).

\begin{figure}
\epsscale{1.0}
\plotone{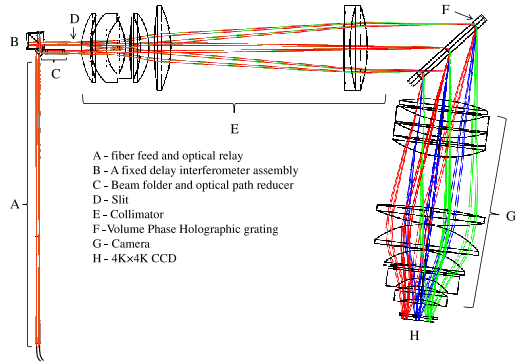}
\caption{
\label{fig:marvels_et1}
Schematic of the MARVELS ET1 instrument.
The ``slit'' is a pseudo-slit containing 60 aligned fibers.
}
\end{figure}

Figure~\ref{fig:marvels_et1} shows the optical layout of the 
MARVELS ET1 instrument \citep{ge09}.
Stellar light from 60 fibers, each subtending
$1\farcs 8$ on the sky, is fed through an optical relay to a
fixed-delay interferometer.
The interferometer creates interference fringes
in each stellar beam. The two outputs of the interferometer from each
input stellar beam are imaged to a slit of an optical spectrograph with
resolution 
$R=11,000$. A total of 120 stellar fringing spectra are formed on a 
4k$\times$4k CCD
detector. Each stellar fringing spectrum covers roughly 
24$\times$4096 pixels (24 pixels along the slit direction and 4096 pixels 
in the dispersion direction). 
The wavelength coverage per spectrum is 
$\lambda \approx 5000-5700\,$\AA.
Environmental stabilization
keeps temperature drifts below $\sim 5\,$mK
during a typical night.  The corresponding radial velocity
drift is less than $20\msec$ within a day without any RV
drift calibration.
Because of this stability, no
iodine cell is needed in the stellar beam path during the science
exposures.  Instead, spectra of a ThAr emission lamp and
an iodine absorption cell illuminated by a tungsten continuum lamp
are taken before and after each science exposure,
and these are used to remove
instrumental drifts.  
In November 2010, at the end of the first 2-year observing
cycle, we replaced the original MARVELS plugplate fibers with
new fibers that subtend $2\farcs 54$ on the sky, which increases
the overall throughput.

\begin{figure}
\epsscale{1.0}
\plotone{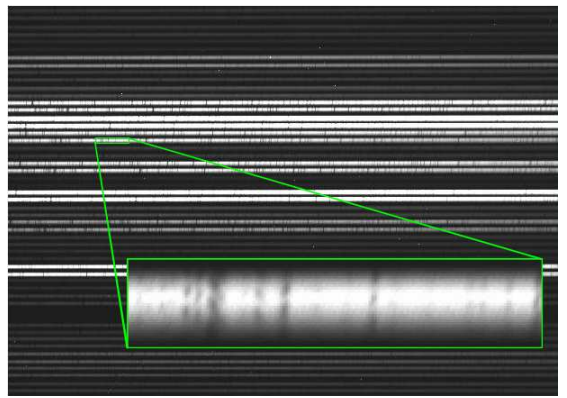}
\caption{
\label{fig:marvels_spectra}
Selected region of a MARVELS science exposure.
Each horizontal stripe represents one of the two interferometer
beam outputs for one of the 60 targets.
The expanded region shows a section of the fringing spectrum of
HIP 14810 \citep{wright09}, a G5 dwarf with
$V=8.5$, over the wavelength range 
$5293\mangstrom < \lambda < 5316\mangstrom$ (150 wavelength channels
out of 4096 in the spectrum).  Each spectral line is broken into a series of
dark features that represent the minima of the sinusoidal fringe pattern.
The shifts in the vertical position of these fringes, fitted over
all lines in the spectrum, are measured to extract
precision radial velocities.
}
\end{figure}

Figure~\ref{fig:marvels_spectra} shows an area selected from
a MARVELS science exposure, with an expanded region that 
shows a portion of the fringing spectrum of an individual 
object, in this case a $V=8.5$ star.
This region can be compared to the bottom panel of
Figure~1 from \cite{vaneyken10}, which shows an idealized 
case of such a fringing spectrum.  The
horizontal axis is the wavelength direction, and each
spectral line produces a sinusoidally
modulated fringe pattern in the vertical direction.
Small line shifts in the wavelength direction produce 
fringe shifts roughly four times larger in the vertical direction.
A $30\msec$ RV change shifts the vertical position of
the fringes by $\sim 0.01$ pixel, and it
is this mean vertical shift that must be measured by
the data pipeline to extract radial velocities (after
removing the much larger but computable effects from
the Earth's rotation and orbital motion).

MARVELS aims
to survey $\sim 8400$ stars in the apparent magnitude range $V=8-12$,
visiting each star approximately 24 times over a two to four year
interval. MARVELS began operations in Fall 2008 with a 60-fiber
instrument known as the W.\ M.\ Keck Exoplanet Tracker (ET).  We hope to
augment the survey with a second, similar instrument (ET2) by Fall
2011, but we have not yet finalized the funding required to do so.
The principal high-level goal of MARVELS is to produce a statistically
well defined sample of $\sim 100$ giant planets with periods up to
two years, drawn from a large sample of host stars that have
well understood selection biases and encompass a wide range of stellar
properties. This data set will be
suitable for revealing the diversity in giant exoplanet
populations and for testing models of the formation, migration, and
dynamical evolution of giant planet systems.  
In addition, the large stellar sample
of MARVELS makes it sensitive to populations of rare systems, which
are often signposts of the physical processes at work in planet
formation or migration, including very hot Jupiters ($P < 3$ days),
short-period super-massive planets ($P < 10$ days, $M \sim 5-15~M_J$),
short-period eccentric planets, planets in extremely eccentric orbits,
planets orbiting low metallicity stars, and rapidly interacting
multiple planet systems.  The systems in which MARVELS identifies
giant planets are ideal targets for systematic follow-up campaigns at
higher RV precision to quantify the frequency of lower mass or longer
period companions in multiple planet systems.  Finally, the large size
and homogeneity of the target sample make MARVELS an ideal experiment
for quantifying the
emptiness of the ``brown dwarf desert'' at masses $M \sim 13-80 M_{\rm
Jup}$ \citep{grether06}
and a unique resource for studying short and intermediate
period binary star populations.

\begin{deluxetable}{l}
\tablecolumns{1}
\tablewidth{0pc}
\tablecaption{Summary of MARVELS\label{tbl:marvels}}
\tablehead{}
\startdata
{\it Duration:} Fall 2008 - Summer 2014, bright time\\
{\it Spectra:} Dispersed fixed-delay interferometer spectrograph\\
\phantom{\it Spectra:}  60 fibers per plate \\
\phantom{\it Spectra: xxxx}  (may increase to 120 from Fall 2011)\\
\phantom{\it Spectra:}  5000\AA\ $< \lambda <$ 5700\AA \\
\phantom{\it Spectra:}  $R \equiv \lambda/\Delta\lambda \approx 11,000$\\
{\it Targets:} 8400 FGK stars, $8 \leq V \leq 12.5$\\
\phantom{{\it Targets:}} 10\% giants\\
\phantom{{\it Targets:}} 24 epochs per star, spread over 2-4 years\\
{\it RV Precision:} $10.5\msec$ $(V \leq 9)$ \\
\phantom{{\it RV Precision:}} $22\msec$ $(V=10)$ \\
\phantom{{\it RV Precision:}} $35\msec$ $(V=11)$ \\
\phantom{{\it RV Precision:}} $45\msec$ $(V=11.5)$ \\
\enddata
\tablecomments{
Number of targets assumes 120 fibers from Fall 2011,
which will increase the magnitude limit from the current 
$V\approx 12$ to $V\approx 12.5$.
Quoted precision goals are $1.3 \times$ median photon noise
from Years 1+2.
}
\end{deluxetable}

\begin{figure}
\epsscale{1.2}
\plotone{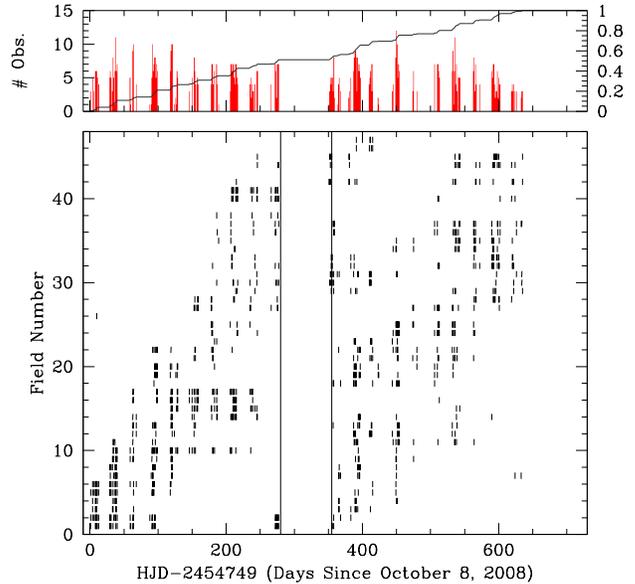}
\caption{
\label{fig:marvels_obs}
({\it Bottom})
The distribution of the epochs (Heliocentric Julian Date)
for each of the 47 MARVELS target
fields during Years 1 and 2. 
Each vertical bar represents an observation.
Vertical lines mark the summer shutdown.
({\it Top})
The histogram shows the number
of observations as a function of HJD (left axis), and
the line shows the cumulative fraction of the total number of
observations (right axis).
}
\end{figure}

During the first two seasons of operation
and the first four months of the third season
(through December 2010), MARVELS
targeted 2820 stars in 47 fields chosen to allow good time coverage
across the sky, to
have sufficient numbers of stars in the $8 \leq V \leq 12$ magnitude
range, and to have several fields within the {\it Kepler}
\citep{batalha06} survey footprint (see Figure~\ref{fig:marvels_obs}).
The median number of epochs
for these fields was 26, with a subset of 2580 stars in 43
fields having at least eighteen observation epochs.
Figure \ref{fig:marvels_obs} shows the distribution of
observation epochs
for each of the target fields.
For the remainder of Year 3 and through Year 6, we
have selected 58 fields for co-observation with APOGEE; we will not
revisit the Year 1+2 fields.

Stars are selected from cross-matched
combinations of the NOMAD \citep{zacharias04}, UCAC3
\citep{zacharias10}, GSC2.3 \citep{lasker08}, and 2MASS
\citep{skrutskie06} catalogs.  Giant stars ($\log g < 3.5$) are
separated from dwarfs and sub-giants using a reduced proper motion
(RPM) diagram, RPM$_J$ vs. $J-H$, using the separation criterion from
\cite{collier07}.  We select the six brightest giants in each field
with $4300\K < \teff < 5100\K$, corresponding to spectral types K2 to
G5.  We exclude dwarfs with $\teff > 6250\K$, which are generally
rotating too rapidly and have too few lines to measure precise
radial velocities using our
instrumentation.  Because the brightest dwarfs in a magnitude-selected sample
are predominantly earlier spectral types, we require that no more than
40\% of our dwarfs in a field have $5800\K < \teff < 6250\K$,
corresponding to spectral types G0 to F7.  We populate our target list
in a given $7\,$deg$^2$ field
by adding the brightest dwarfs until we have
22 from the G0 to F7 set, then continue in order of decreasing
brightness but selecting only those dwarfs with $\teff < 5800\K$.
Note that for the fields targeted in Years 1+2, we used somewhat
different target selection criteria, and also used spectroscopic
observations (with the SDSS spectrographs) for giant-dwarf separation.
Unfortunately, the SDSS spectra (reduced with an earlier version of
the SSPP) proved less effective in separating giants and dwarfs
than we had expected, leaving us with a 30\%
giant fraction compared to our original goal of 10\%.
The RPM$_J$ selection we have now implemented should resolve this
problem.

MARVELS observes with 50-minute science exposures (which
will increase to 60 minutes once co-observing with APOGEE begins)
and $\sim 10$ minutes of overhead 
per exposure.  The fields for any given night are selected based on
observability, the number of previous epochs, and the time since the most
recent epoch.
The photon-noise limited RV precision for a MARVELS observation
depends most strongly on stellar apparent magnitude, but also
on other factors that affect fringe visibility including
rotation and metallicity.  
For observations during the first
two years, the median RV photon-noise limited precision is
approximately $5\msec$ at $V=8.5$, $8\msec$ at $V=9.0$,
$17\msec$ at $V=10.0$, $27\msec$ at $V=11.0$, and
$35\msec$ at $V=11.5$.  The median rms RV of target stars
observed over a one-month timescale 
is $2-3$ times the photon noise at bright magnitudes
and $1-1.5$ times the photon noise at $V \geq 10$.
The current MARVELS data pipeline has shortcomings that lead to
worse performance over timescales of several months.
We are working on improvements as of this writing,
with the eventual goal of achieving total errors
within a factor 1.3 of the photon noise errors.
The precision goals in Table~\ref{tbl:marvels} are defined
by this target, using the photon-noise numbers quoted above.
We are also implementing changes to the fiber system and data pipeline
that we expect to lower the photon noise itself by a moderate factor
(e.g., changing from iodine to ThAr calibration, 
which will allow us to use 100\% of
the spectral range rather than the current 75\%).
We therefore regard it as plausible that the final MARVELS
RV performance will improve on the goals in Table~\ref{tbl:marvels}
by $\sim 30\%$.

The velocity semi-amplitude of a star of mass $M_*$
orbited by a planet of mass $M_p$ with a period $P$ and
inclination $i$ is
\begin{equation}
K = 28.4 \msec (M_p \sin i /M_{\rm Jup})(P/1\,{\rm yr})^{-1/3}
(M_*/M_\odot)^{-2/3} .
\end{equation}
For $N$ observations and an rms radial velocity error of $\sigma$,
achieving less than one false detection for $\sim 10^4$ stars
requires a total signal-to-noise ratio threshold
of approximately $\sqrt{N/2}(K/\sigma) \sim 13$	
\citep{cumming04}, or $\sigma \la K/4$
assuming 24 observations per star.
A simple and somewhat conservative forecast, described in 
Appendix B,
indicates that MARVELS should detect approximately 66 planets with
$P<2$ year and $M_p< 10~M_{\rm Jup}$ if the total errors can be
reduced to 1.3 times the median photon noise achieved in Year 1+2 data.  
Of these 66 predicted planets, 53 have
periods of $<1$ year.  The yield falls to 41 planets if the errors are
2.0 times the median photon noise, and it rises to $\sim 86$ planets if
the errors are equal to the median photon noise.  
With the above mentioned changes to the fiber system and data
processing techniques, we may be able to significantly lower
the photon noise floor, which could increase the yields by as much
as 30\%.


These forecasts
assume a second MARVELS instrument operating for the final three years
of the survey; without it, anticipated planet yields fall by $\sim
20\%$ (not 50\% because the second instrument
would be observing fainter stars than the first one).
The predictions are based on a false alarm probability of $\sim
3\times 10^{-4}$, for which we would expect $\sim 3$ false positives.
For a more conservative false alarm probability of $\sim 3\times
10^{-5}$, the yields decrease by $\sim 15\%$.  We note that
we have not attempted to estimate the planet yield from our sample of giant
stars, which constitute $\sim 10\%$ of our targets, as considerably
less is known about the frequency of planets around these systems.
Nevertheless, we can reasonably expect to detect additional planets
from this sample. Finally, we have only included companions with
masses of $< 10\,M_{\rm Jup}$ in our tally.  Extrapolating the planet
distribution function found by \cite{cumming08} up to larger masses,
we estimate that we would detect an additional $\sim 14$ planets with
$10~M_{\rm Jup}< M_p < 15~M_{\rm Jup}$ and periods of $<2$ years under
the assumption of errors equal to $1.3\times$ the photon noise.
Because the observing strategy, target selection, and noise
characteristics for MARVELS are very well specified, statistical
models of the planet population (specifying, e.g., the distribution
of masses, periods, and eccentricities as a function of host properties)
can be tested statistically against the MARVELS RV measurements 
even without one-by-one identification of planets.

\begin{figure}
\epsscale{1.0}
\plotone{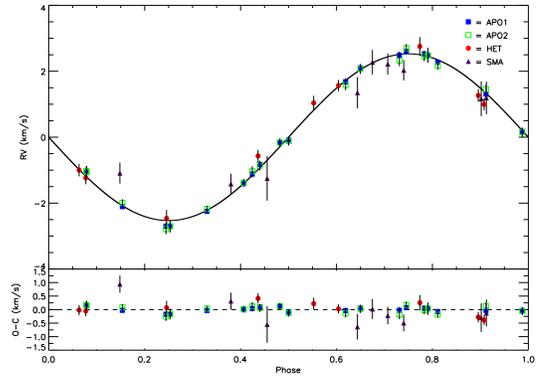}
\caption{
\label{fig:marvels_bd}
The radial velocity curve of the first MARVELS RV companion,
a probable brown-dwarf in a 5.9-day orbit
around the star TYC 1240-00945-1 \citep{lee10}.
Filled and open squares are measurements from the 
two interferometer outputs of the MARVELS spectrograph.
Circles and triangles are measurements from follow-up observations
from the Hobby-Eberly Telescope and the SMARTS 1.5m telescope, respectively.
Lower panel shows the difference between observed velocities and those
calculated from the best-fit model.
}
\end{figure}

Figure~\ref{fig:marvels_bd} shows the MARVELS RV curve for 
the short-period brown dwarf candidate discovered by MARVELS around
the star TYC-1240-00945-1 \citep{lee10a}.
RV measurements from the two interferometer beams are shown separately,
as well as measurements from observations with
the Hobby-Eberly Telescope (HET) and the SMARTS 1.5m echelle. 
Supplementary photometric and spectroscopic studies show that 
the host is a slightly evolved,  
solar-type star, with an estimated mass of $1.35 M_\odot$ and
age of $\approx 3.0$ Gyr.  In this case, the low-mass companion
(``MARVELS-1b'') is a likely brown dwarf with minimum
mass of $28.0 \pm 1.5 \,M_{\rm Jup}$ at an orbital separation of 
$0.071 \pm 0.002$ AU ($P_{\rm orb} = 5.9$ days), placing this object
squarely within the ``brown dwarf desert". Indeed, MARVELS
has already found several more such brown-dwarf ``desert dwellers" in
the first two-year dataset, 
which will enable us to quantify the aridity of the desert
(N.\ De Lee et al.\ 2011, in preparation).
The MARVELS team plans similarly detailed follow-up studies and
characterization of all MARVELS hosts and control samples of target
stars that did not yield planets.  This approach will enable full 
investigation of the dependence of giant exoplanet and brown dwarf
populations on host star properties, including chemical abundances,
mass, and evolutionary status.
In addition, MARVELS will provide robust
statistics on spectroscopic binary star populations, 
and it will yield a novel sample of 
eclipsing binary star systems {\it discovered} spectroscopically.

\section{Science Organization}
\label{sec:org}

An effective collaboration culture is crucial to the successful execution
of a large project like the SDSS.
Indeed, developing this culture was itself one of the major challenges
and significant achievements of SDSS-I.
The organization of SDSS-III is, of course, closely modeled on that  of
SDSS-I and II.  We briefly describe this organization here, as it may
be of value to those using the SDSS-III data
sets and science analyses and to others planning comparably ambitious
projects.

Like its predecessors, SDSS-III is being carried out by a large and
diverse
international collaboration. A wide variety of institutions have joined
the project by means of financial or
equivalent in-kind contributions, and they all agree to a written set of
``Principles of Operation"\footnote{Available
at {\tt http://www.sdss3.org/collaboration/poo3.pdf}} that serves
as the defining policy document of the project.
At {\it Full Member} institutions, all faculty, PhD research staff,
and students have access rights to all SDSS-III data and activities.
{\it Associate Member} institutions join with smaller, designated
groups of faculty and postdoctoral researchers.  A {\it Participation Group} 
is a consortium of designated researchers 
from multiple institutions that acts as a single member institution
within the SDSS Collaboration.
Finally, particular individuals are named as {\it External Participants}
based on their contributions to the SDSS-III project.
An up--to--date listing of all the institutions in SDSS-III can be found
on the SDSS-III website ({\tt http://www.sdss3.org}).
The Apache Point Observatory and the Sloan Foundation Telescope are both
owned by the Astrophysical Research Consortium (ARC), and the ARC Board of
Governors has financial authority for the SDSS-III.  
An Advisory Council oversees the survey and represents 
the collaboration to the ARC Board of Governors.
The Advisory Council consists of one voting member from each full member,
participation group, and associate member of sufficient group size.

\begin{figure*}
\epsscale{0.9}
\includegraphics[clip=true,scale=0.7,angle=90]{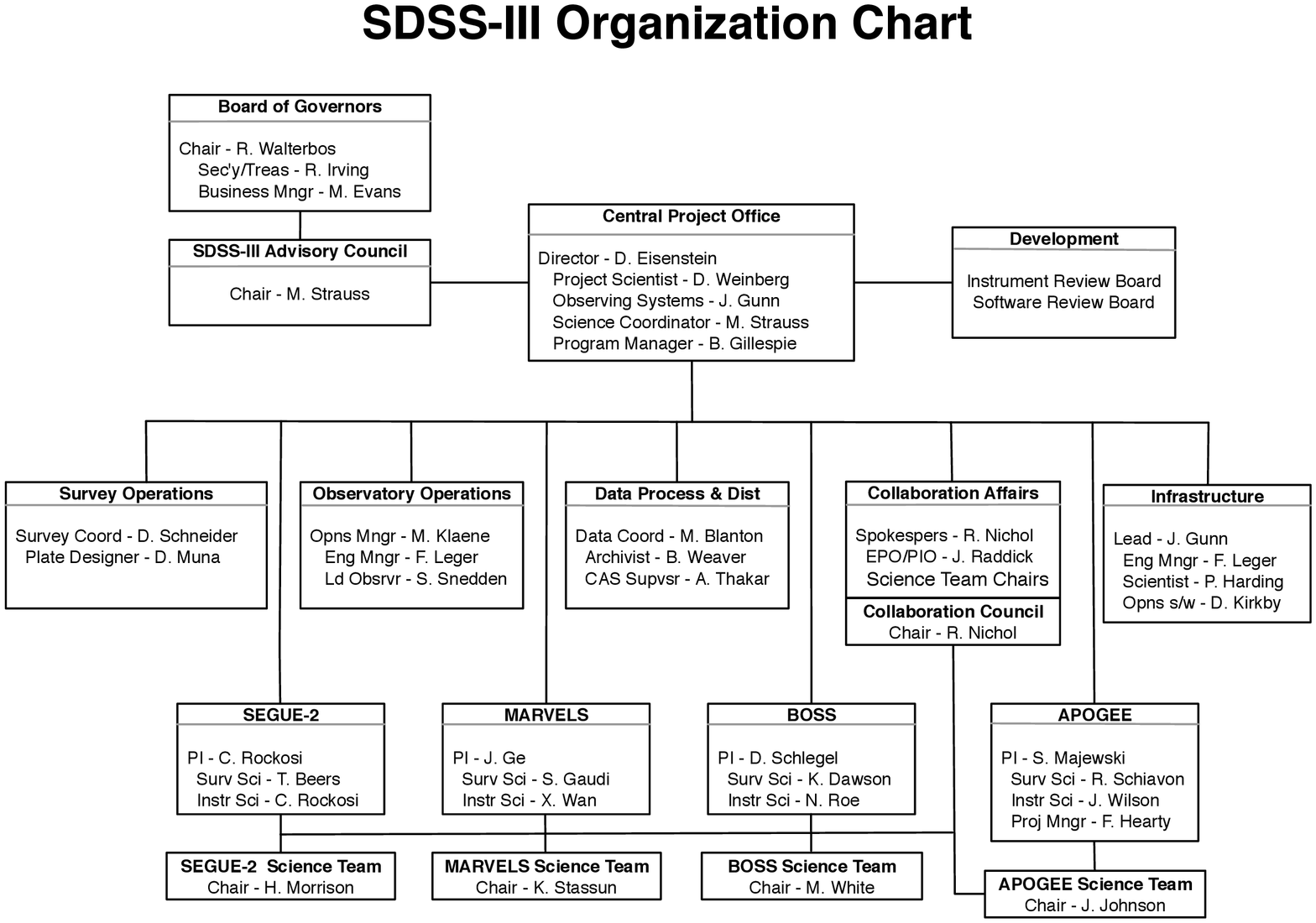}
\caption{
\label{fig:orgchart}
The high-level organizational chart for SDSS-III.
Named individuals are those filling these positions as of January 1, 2011.
As is evident from the author list of this paper,
this chart represents the tip of a very large iceberg of SDSS-III 
contributors.
}
\end{figure*}

Figure~\ref{fig:orgchart} presents the high-level SDSS-III organization
chart, including the individuals who currently hold the indicated
positions.  Each of the four surveys has its own technical and science
team;
the number of people who have already made large contributions (i.e.,
many person-months or more)
to the design and execution of the individual surveys ranges from
$\sim 15$
to $\sim 60$, with still larger numbers joining in data analysis and
quality assurance.
The Principal Investigator (PI) of each survey oversees all aspects of
the survey's
construction and execution, with the assistance of the Survey Scientist,
who has the primary responsibility for defining science requirements
and ensuring that the survey data ultimately meet those requirements.
BOSS, MARVELS, and APOGEE each have an Instrument Scientist
who oversees the design, construction, commissioning, and maintenance of
the new instrumentation.  The APOGEE spectrograph development
has its own Project Manager, as did the MARVELS ET1 spectrograph
development prior to delivery of the instrument.

Many tasks, including overall project budgeting and management,
span all four surveys of SDSS-III.  Organizational responsibility
for these tasks lies with the central project office, headed
by the Director.  The Infrastructure and Observatory Operations teams have
responsibility for common facilities (telescope, fiber systems,
operations software, etc.) and for performing the 
observations themselves.
The Data Coordinator
is responsible for integrating data from the four surveys into the
science data archive, the basis
both for collaboration science and for public data releases.
The Project Spokesperson, elected by the collaboration,
promotes scientific coordination within the collaboration and 
external visibility of SDSS-III in the astronomical community and beyond.
The Spokesperson chairs the Collaboration Council, comprised of
representatives
from all voting institutions, which organizes collaboration meetings
and develops and implements collaboration policies, most notably
on publications and external collaborators.\footnote{See
{\tt http://www.sdss3.org/collaboration/policies.php}.}
Over the years, collaborations with non-participants on 
pre-publication data have been a vital mechanism for bringing
additional expertise and resources into SDSS science analyses,
and reviewing external collaborator proposals is one of the
Collaboration Council's most important tasks.

The guiding principle of the SDSS-III science collaboration
is that all participants have the 
right to pursue any project they wish with SDSS-III data, 
but they are required to notify the entire collaboration
of their plans and to update them as projects progress.
Groups pursuing similar science projects
are encouraged to collaborate, but they are not required to do so. There
is no binding internal refereeing process, but draft publications
using non-public data must be posted to the whole collaboration for a
review period of at least three weeks prior to submission to any journal
or online archive.  Manuscripts often undergo significant revision
and improvement during this period. 
Participants
outside of the core analysis team may request co-authorship on a
paper if they played a significant role in producing the data or
analysis tools that enabled it. In particular,
scientists who have contributed at least one year of effort to SDSS-III
infrastructure development or operations can request ``Architect''
status, which entitles them
to request co-authorship on any science publications for those surveys
to which they contributed. 
All SDSS-III authorship requests
are expected to comply with the professional guidelines 
of the American Physical 
Society.\footnote{{\tt http://www.aps.org/policy/statements/02\_2.cfm}}

Each of the four SDSS-III surveys has its own survey science team (SST),
headed by the SST Chair, whose role is to coordinate and promote effective
scientific collaboration within the team.
Naturally, the science team overlaps and interacts with the survey's technical
team, but the latter is focused on producing the data, while the former
is focused on science analysis;
data quality assurance is a shared responsibility. Inside an SST,
participants may coordinate their efforts on more focused topics
via working groups; for example, the BOSS SST presently has working
groups in galaxy clustering, galaxy evolution, \lya\ forest cosmology,
physics of the intergalactic medium, and quasars. 
The working groups communicate and collaborate through archived
e-mail lists, wiki pages, regular teleconferences, and in-person meetings.
As a result, many SDSS papers benefit from the combined
efforts and knowledge of many collaborators on the science analysis,
as well as on the production of the data that enables it.

Over the years, all of these strategies --- representative governing
bodies, centralized management overseeing the many technical teams,
well defined policies and structures that encourage widespread
scientific participation, and extensive communication mechanisms ---
have proven essential to the successful execution of the SDSS and
to producing an enormous range of science from its surveys.

\section{Concluding Remarks}
\label{sec:summary}

The SDSS has demonstrated the extraordinary scientific reach of
a moderate aperture telescope equipped with powerful wide-field
instruments and operated in an efficient ``survey mode,'' including
sophisticated data pipelines that produce well calibrated
and readily usable public data sets.
There are now several astronomical imaging cameras in operation or under
construction that exceed the pixel count of the SDSS imager,
which has been officially retired after completing the additional
southern imaging for BOSS.  
However, the Sloan Telescope remains an exceptionally 
productive facility for wide-field spectroscopic surveys.

The four SDSS-III surveys exploit this capability efficiently to
address a wide range of science goals.  
The BOSS spectroscopic survey requires five years
of dark time to cover its 10,000 deg$^2$ survey area.
SEGUE-2 and the BOSS imaging survey used the one year of dark time 
that was available between the end of SDSS-II and the completion
of the BOSS spectrograph upgrades.
MARVELS began bright-time operations in the first fall season of
SDSS-III.  MARVELS and APOGEE will share the focal plane for 75\%
of the bright-time observing from 2011-2014, allowing each survey
to amass a considerably larger sample than it could with a 50\% share.

The BOSS galaxy redshift survey will achieve BAO distance scale
constraints that are close to the limit set by cosmic variance
out to $z=0.6$; the only substantial (factor-of-two) improvement
possible at these redshifts
would be to cover the remaining $3\pi$ steradians of the sky.
The BOSS \lya\ forest survey will pioneer a new method of 
measuring three-dimensional structure in the high-redshift universe
and provide the first BAO measurements of distance and expansion
rate at $z>2$.  Together SEGUE and APOGEE will provide
powerful new insights into
the formation history and present day structure of the Milky Way.
The depth and large sample size of the SEGUE surveys make them
especially valuable for characterizing kinematic and chemical
structure in the outer Galaxy.  The dust-penetrating capacity of APOGEE's
infrared observations will make it the first large spectroscopic survey of 
all Galactic stellar populations, and its high resolution and
high precision allow detailed chemical fingerprinting of an
enormous sample, orders of magnitude larger than any 
high resolution sample that exists today.
The large sample of stars monitored by the MARVELS RV survey
gives it sensitivity to rare planetary systems that are
signposts for underlying physical processes, and the combination
of sample size and systematic observing strategy will make it
a uniquely valuable data set for testing theories of giant planet
formation, migration, and dynamical interaction.

Current investigations with SDSS-III data span a vast range of 
scales and redshifts, including studies of large-scale
structure with massive galaxies 
and three-dimensional \lya\ forest correlations,
searches for kinematic and chemical substructure in the Galactic halo
and thick disk, and measurement of the incidence of short-period
brown dwarf companions to solar-type stars.
SDSS-III will continue the long-standing SDSS tradition of 
public data releases, beginning with the SDSS Eighth Data Release (DR8),
which is now available \citep{dr8}.  
DR8 includes all of the new imaging carried
out for BOSS and all of the spectra taken for SEGUE-2.  It also 
incorporates all SDSS-I and II data, processed with the latest
versions of our data reduction and calibration pipelines, 
so that science analyses can incorporate data from all SDSS surveys
in a seamless and internally consistent manner.  The final data
SDSS-III release is scheduled for the end of 2014.  Like their
predecessors, we anticipate that BOSS, SEGUE-2, MARVELS, and APOGEE
will have deep and wide-ranging impacts on many fields of
contemporary astronomy and cosmology.

\acknowledgments

Funding for SDSS-III has been provided by the Alfred P. Sloan Foundation,
the Participating Institutions, the National Science Foundation, and the
U.S. Department of Energy Office of Science. 
The SDSS-III web site is http://www.sdss3.org/.

SDSS-III is managed by the Astrophysical Research Consortium for the
Participating Institutions of the SDSS-III Collaboration including the
University of Arizona, the Brazilian Participation Group, Brookhaven
National Laboratory, University of Cambridge, 
Carnegie Mellon University, University of Florida,
the French Participation Group, the German Participation Group, 
Harvard University,
the
Instituto de Astrofisica de Canarias, the Michigan State/Notre Dame/JINA
Participation Group, Johns Hopkins University, Lawrence Berkeley National
Laboratory, Max Planck Institute for Astrophysics, New Mexico State
University, New York University, Ohio State University, Pennsylvania
State University, University of Portsmouth, Princeton University, the
Spanish Participation Group, University of Tokyo, University of Utah,
Vanderbilt University, University of Virginia, University of Washington,
and Yale University.



{\it Facilities:} \facility{Sloan}



\appendix

\section{Forecasts for BOSS}
\label{appx:boss}


In Table \ref{tbl:fom} we forecast the constraints
on a number of cosmological parameters.
To obtain these numbers we first convert our observational parameters into
errors on the line-of-sight [$H(z)$] and transverse [$d_A(z)$] distances as a
function of redshift using the method of \citet{seo07}.
This Fisher matrix calculation uses only acoustic oscillation information and
no broad-band power, so we believe the error estimates to be robust (and
conservative, see the discussion below and the final line
of Table~\ref{tbl:fom}).
To approximate the effects of (partial) reconstruction \citep{eisenstein07a}
we suppress the nonlinear smearing ($\Sigma_{\perp,||}$ in the notation of
\citealt{eisenstein07b}) by a factor of two for the LRG calculation.
We use a similar Fisher matrix calculation \citep{mcdonald07} to estimate
the distance errors that one would obtain for the \lya\ forest survey,
with no attempt at reconstruction because of the very sparse sampling
of the density field.
We find errors on $d_A$ of 1.0\% at $z=0.35$, 1.0\% at $z=0.6$,
and 4.5\% at $z=2.5$,
with errors on $H(z)$ of 1.8\%, 1.7\% and 2.6\% at the same redshifts.
As noted earlier, current theoretical studies suggest that any shifts in
the BAO scale due to non-linearity or galaxy bias are at or below this
level.  With further work, we should be able to calculate any corrections
to a level of accuracy that keeps systematic errors well below these
statistical errors.

The constraints on $d_A$ and $H$ are then used in a Fisher matrix
calculation to get
constraints on the matter density $\omega_m\equiv\Omega_mh^2$, the baryon
density $\omega_b\equiv\Omega_bh^2$, the dark energy density $\Omega_{X}$,
$w_0$, $w_a$, and the curvature, $\Omega_K$. 
The dark energy
equation-of-state parameter is assumed to evolve with expansion factor
$a(t)$ as $w(a) = w_0 + w_a(1-a) = w_p + w_a(a_p-a)$, where the
``pivot'' expansion factor is the one at which errors on $w_p$
and $w_a$ are uncorrelated.
In addition to the distance constraints from BAO experiments,
we add the Fisher matrices for Planck and Stage II experiments
presented in the technical appendix of the 
Dark Energy Task Force (DETF) report \citep{albrecht06}.
The variance of each parameter is given in
Table \ref{tbl:fom}.
We also quote the DETF Figure of Merit, which is the inverse of the area of
the 95\% confidence level region in the $w_p-w_a$ plane
(scaled to correspond to the convention adopted by the DETF).
The precise value of the pivot expansion factor depends on which experiments
are considered, but it is generally $a_p \approx 0.75$, i.e., in this
family of models the dark energy experiments best constrain the value
of $w$ at $z \approx 0.35$.

\begin{deluxetable}{lcccccc}
\tablecolumns{7}
\tablewidth{0pc}
\tablecaption{BOSS Parameter Constraint Forececasts\label{tbl:fom}}
\tablehead{Expt. & $h$ & $\Omega_K$ & $w_0$ & $w_p$ & $w_a$ & FoM}
\startdata
Planck+Stage II & 0.019 & 0.0031 & 0.115 & 0.036 & 0.524 &  53.4 \\
BOSS LRG BAO  & 0.009 & 0.0027 & 0.090 & 0.031 & 0.365 &  87.4 \\
{\bf BOSS LRG BAO+LyaF BAO} & 0.009 & 0.0019 & 0.083 & 0.030 & 0.320 & {\bf
102}
\\
BOSS LRG broad-band+LyaFBAO & 0.007 & 0.0018 & 0.074 & 0.019 & 0.284 & 188
\enddata
\tablecomments{
All constraints assume Planck and the DETF forecasts for ``Stage II''
experiments.  BAO constraints include only the acoustic scale information
and
are therefore conservative; the final line shows the BOSS forecast that
also incorporates broad-band galaxy power information.
}
\end{deluxetable}

\section{Forecasts for MARVELS}
\label{appx:marvels}

To estimate the MARVELS survey planet yield, we use a simple,
essentially analytic, method. We adopt the following assumptions:

\noindent (1) We consider $9$ bins in $V$ magnitude, from $V_i=8$ to
$V_i=12$ in steps of 0.5 magnitude, where $i$ is the bin index. For
stars monitored in Years 1+2 including the 4 month extension in Year
3, we use the actual distribution of the number of stars per bin
$f_{*,i}$ from a representative subset of the target fields.  For
stars monitored in Years 3-6, we use an average distribution of $V$
magnitudes for dwarf stars contained in the 31 preliminary shared
fields chosen by APOGEE.  We assume that the current MARVELS spectrograph
monitors the brightest 60 stars and the second instrument 
(assumed to begin operation in Fall 2011) monitors the next brightest 60 stars.

\noindent (2) The total number of stars $N_*$ monitored is set by the
number of observations $N_{\rm obs}$ per field, the total number of epochs
$N_e$ available per month, the total number of available months $N_m$
for each observing block, and the fraction of time $f_{\rm lost}$ lost to
non-MARVELS science (APOGEE commissioning, APOGEE-only fields, etc.),
\begin{equation}
N_* = N_{\rm fiber} \times \frac{N_e N_m}{N_{\rm obs}} \times 
  \left(1-f_{\rm lost}\right).
\label{eqn:nstar}
\end{equation}
We assume a total of 46 epochs are available per month, based on the
917 epochs obtained over the 20 months of Years 1+2.  This
effectively means that for Years 3-6 we assume similar weather and a
similar exposure plus overhead time ($\sim 50+10$ minutes) for each
observation as we adopted in Years 1+2.
We take $f_{\rm lost} = 0.25$ starting in 2011 and $f_{\rm lost}=0$
before.

\noindent (3) We assume a log-normal distribution of RV uncertainties
$\sigma$ for each magnitude bin,
\begin{equation}
\frac{df_{\sigma,i}}{d\sigma} \propto e^{-0.5[
(\log{\sigma}-\log{\sigma_i})/0.2]^2},
\label{eqn:dndsig}
\end{equation}
where the $\sigma_i$ are the median photon noise uncertainties
and RMS scatter for each $V$ magnitude bin.  The value of 0.2
for the log-normal dispersion was chosen to approximately match the
distribution of observed scatter.  Our
final yields do not depend very strongly on this choice, changing by
$\sim 27\%$ over the range $0.05-0.4$.

\noindent (4)
We adopt a power-law distribution for the planet frequency as a
function of period and mass,
\begin{equation}
\frac{dN_p}{d\ln{m_p}d\ln{P}} = C(P) \left(\frac{m_p}{M_{\rm
Jup}}\right)^{-0.31},
\label{eqn:dndmdp}
\end{equation}
where $C(P) \simeq 0.00186$ for $P<300$ days and $C(P)=0.0093$ for $P>300$
days.
We assume that there are no planets with $P<10$ days and $m_p > 2M_{\rm Jup}$.
This distribution is motivated by the results of \cite{cumming08},
who fit a continuous power-law over the entire period
range of $P<2000$ days, finding $dN_p/{d\ln{m_p}d\ln{P}} \propto
m_p^{-0.31}P^{0.26}$.  
However, because they have a paucity of intermediate period planets
($P=10-300$ days) relative to this fit, we have conservatively
chosen a uniform distribution in $\ln{P}$ and step-function 
truncations.  Adopting the continuous power-law distribution
would have led to a higher predicted yield.

\noindent (5)
We assume a planet is detectable if its periodogram power $z$, or
equivalently total signal-to-noise ratio $Q$,
is larger than a given value $z_0$.  We use an analytic estimate for $Q$
as a function of $N_{obs}$, the planet semiamplitude $K$, and the RV
uncertainty given by
\begin{equation}
Q \equiv \left(\frac{N_{\rm obs}}{2}\right)^{1/2}\frac{K}{\rm \sigma}.
\label{eqn:q2}
\end{equation}
This is strictly only appropriate for a uniformly-sampled, circular orbit,
but it is
a good approximation for eccentricities less than $\sim 0.6$, and for
planet periods less than the time spanned
by the observations.   The relationship between the periodogram power $z$
and $Q$ for a Keplerian orbit
fit is given by \citep{cumming04},
\begin{equation}
z \equiv \frac{Q^2}{2}\left(\frac{N_{\rm obs}-5}{N_{\rm obs}}\right),
\label{eqn:z}
\end{equation}
where $N_{\rm obs}-5$ is the number of degrees of freedom from a
Keplerian fit.	We use the following
simple analytic estimate for the minimum power $z_0$ for detection
\citep{cumming04},
\begin{equation}
z_0 = \frac{3\nu}{4} \left[\left(\frac{M}{F}\right)^{2/\nu} -1\right],
\label{eqn:z0}
\end{equation}
where $\nu \equiv N_{\rm obs}-5$,
$M \simeq T\Delta f$ is the number of independent frequencies
(periods) searched for
planets, $T=2$ years is the span of each set of observations, $\Delta f
\equiv P_{\rm min}^{-1}-P_{\rm max}^{-1}$
is the range of frequencies searched, and $F$ is the false alarm probability
required for detection.	
We set $F=(3\times 10^{3})^{-1}$.
Given that MARVELS will be surveying ${\cal O}(10^4)$ stars, we therefore
expect $\sim 3$ false positives, which will require follow-up to
eliminate.  

With these assumptions, we can estimate our yield as,
\begin{equation}
N_{\rm det} =  N_* \sum_{i=1}^9 f_{*,i} 
               \int_{\ln{P_{\rm min}}}^{\ln{P_{\rm max}}}
d\ln{P} \int_{\ln{m_{p,{\rm min}}}}^{\ln{m_{p,{\rm max}}}} d\ln{m_p}
\frac{dN_p}{d\ln{m_p}d\ln{P}} \int_0^{\infty} d\sigma
\frac{df_{\sigma,i}}{d\sigma} \Theta[z(m_p,P,\sigma)-z_0],
\label{eqn:ndet}
\end{equation}
where $\Theta[x]$ is the Heaviside step function. In addition, we exclude
from the integrals
the region where $m_p>M_{\rm Jup}$ and $P<10$ days.





\end{document}